\documentclass[a4paper,11pt]{article}
\pdfoutput=1 

\usepackage{jheppub} 

\usepackage{amsfonts}
\usepackage{mathrsfs}
\usepackage{amsmath}
\usepackage{xcolor}
\usepackage{amssymb}
\usepackage{bm}
\usepackage{stmaryrd}
\usepackage{booktabs}
\usepackage{slashed}
\usepackage{graphicx}
\usepackage{subfigure}
\usepackage{multirow}
\usepackage{verbatim} 
\usepackage[normalem]{ulem}
\allowdisplaybreaks[4]

\usepackage{hyperref}

\definecolor{gesflanse}{rgb}{0.00,0.50,0.50}

\definecolor{gesfpurple}{rgb}{0.47,0.19,0.42}

\definecolor{gesfblue}{rgb}{0.08,0.42,0.76}

\definecolor{gesfred}{rgb}{1,0,0}

\newcommand{\ured}[1]{{\color{gesfred}\underline{#1}}}

\newcommand{\gsec}[1]{{\hypersetup{linkcolor=red}Sec.\,\ref{#1}\hypersetup{linkcolor=blue}}}

\newcommand{\Ssec}[1]{{\hypersetup{linkcolor=red}Section\,\ref{#1}\hypersetup{linkcolor=blue}}}

\newcommand{\gapp}[1]{{\hypersetup{linkcolor=red}Appendix\,\ref{#1}\hypersetup{linkcolor=blue}}}
\newcommand{\geqn}[1]{(\ref{#1})}
\newcommand{\gfig}[1]{{\hypersetup{linkcolor=violet}Fig.\,\ref{#1}\hypersetup{linkcolor=blue}}}
\newcommand{\gtab}[1]{{\hypersetup{linkcolor=gesflanse}Table\,\ref{#1}\hypersetup{linkcolor=blue}}}

\newcommand{\fr}[2]{\mbox{$\frac{{#1}}{#2}$}}

\renewcommand{\rm}{\mathrm}
\def\bge{\begin{equation}}
\def\ede{\end{equation}}
\def\bga{\begin{aligned}}
\def\eda{\end{aligned}}
\newcommand{\beq}{\begin{equation}}
\newcommand{\eeq}{\end{equation}}
\newcommand{\bq}{\begin{equation}}
\newcommand{\eq}{\end{equation}}
\newcommand{\ba}{\begin{array}}
\newcommand{\ea}{\end{array}}
\newcommand{\beqa}{\begin{eqnarray}}
\newcommand{\eeqa}{\end{eqnarray}}
\newcommand{\beqs}{\begin{subequations}}
\newcommand{\eeqs}{\end{subequations}}

\def\dis{\displaystyle}

\def\({\left(}
\def\){\right)}
\def\[{\left[}
\def\]{\right]}

\def\End{\end{document}}

\newcommand{\diag}{\mbox{diag}}

\def\geqq{\geqslant}

\def\dif{\partial}

\def\th{\mathrm{th}}
\def\exp{\mathrm{exp}}
\def\BR{\mathrm{Br}}
\def\sm{\mathrm{sm}}
\def\OO{\mathcal{O}}

\def\to{\rightarrow}

\def\GF{G_F^{}}
\def\MZ{M_Z^{}}
\def\MW{M_W^{}}

\def\LambdaTeV{\Lambda_\rm{\scriptscriptstyle TeV}}

\def\sm{\mathrm{sm}}
\def\inv{\mathrm{inv}}

\def\th{\mathrm{th}}
\def\exp{\mathrm{exp}}

\newcommand{\Br}{\textrm{Br}}

\title{
Probing New Physics Scales from Higgs and
Electroweak Observables at $\boldsymbol{e^+ e^-}$ Higgs Factory}

\author[a]{Shao-Feng Ge,}
\author[b,c]{~Hong-Jian He,}
\author[b,d]{~Rui-Qing Xiao}

\affiliation[a]{Max-Planck-Institut f\"{u}r Kernphysik, Heidelberg 69117, Germany}
\affiliation[b]{Institute of Modern Physics and Center for High Energy Physics,\\
                Tsinghua University, Beijing 100084, China}
\affiliation[c]{Center for High Energy Physics, Peking University, Beijing 100871, China}
\affiliation[d]{Lawrence Berkeley National Laboratory, Berkeley, California 94720, USA}

\emailAdd{gesf02@gmail.com}
\emailAdd{hjhe@tsinghua.edu.cn}
\emailAdd{ruiqingxiao@lbl.gov}  

\abstract{\\
New physics beyond the standard model (SM) can be model-independently formulated
via dimension-6 effective operators,
whose coefficients (cutoffs) characterize the scales of new physics.
We study the probe of new physics scales from the electroweak precision
observables (EWPO) and the Higgs observables (HO)
at the future $e^+e^-$ Higgs factory (such as CEPC).
To optimize constraints of new physics from all available observables,
we establish a scheme-independent approach. With this formulation, we treat the SM
electroweak parameters and the coefficients of dimension-6 operators
on equal footing, which can be fitted simultaneously by
the same $\chi^2$ function. As deviations from the SM are generally small,
we can expand the new physics parameters up to linear order and perform an
analytical $\chi^2$ fit to derive the potential reach of
the new physics scales. We find that the HO from both Higgs produnction and
decay rates can probe the new physics scales up to 10\,TeV\,
(and to 44\,TeV for the case of gluon-involved operator $\mathcal{O}_g^{}$),
and the new physics scales of Yukawa-type operators can be probed
by the precision Higgs coupling measurements up to ($13-25$)\,TeV.
Further including the EWPO can push the limit up to $35$\,TeV.
From this prospect, we demonstrate that the EWPO measured in the early phase
of a Higgs factory can be as important as the Higgs observables.
These indirect probes of new physics scales at the Higgs factory can mainly cover
the energy range to be directly explored by the next generation hadron colliders of
$pp\,(50-\!100$\,TeV), such as the SPPC and FCC-hh.
\\[2mm]
\phantom. \hfill JHEP\,(2016), in Press~$[${arXiv:1603.03385}$]$
}

\begin{document}
\maketitle
\flushbottom
\setcounter{page}{2}

\vspace*{5mm}
\section{Introduction}
\label{sec:1}
\vspace*{1mm}

The LHC discovery\,\cite{Higgs12} of a light Higgs boson $h(125\text{GeV})$ \cite{Higgs}
has completed the particle spectrum of the standard model (SM) of particle physics.
This culminates in the success of searches that lasted for
decades \cite{HiggsProfile}. Although the new physics has not yet been
established so far, there are already strong motivations for going beyond the SM,
including the observed neutrino oscillations and cosmic baryon asymmetry, as well as evidences
for the dark matter and inflation. Since 2012, the particle physics has come to a turning point
at which the precision Higgs measurements have become an important task for seeking
clues to the new physics discovery \cite{Ellis}.

We should stress that the SM is not merely a collection of various observed particles
(fermions and bosons). The completion of the SM particle spectrum does not mean the
completion of the SM itself until {\it all SM interaction forces} could be firmly measured.
In fact, the SM consists of three fundamental gauge forces as its key ingredients,
the electromagnetic force, the weak force, and the strong force, which are
mediated by spin-1 gauge bosons, the photon ($A_\mu^{}$), the weak bosons $(W_\mu^\pm,Z_\mu^0)$,
and the gluons $(G_\mu^a)$, respectively.
Furthermore, the spin-0 Higgs boson $h(125\text{GeV})$ is not merely another particle in the SM,
because it joins three types of fundamental forces:
(i) the gauge forces mediated by the spin-1 weak gauge bosons $(W,Z)$;
(ii) the Yukawa forces with fermions mediated by the spin-0 Higgs boson $h$;
(iii) and the cubic and quartic Higgs self-interactions\footnote{%
Note that only the spin-0 Higgs boson can have strict self-interaction force,
while other particles (such as the spin-1 Non-Abelian gauge bosons or the spin-2 gravitons)
cannot, because any interaction vertex of these spin-1 or spin-2 particles always involves
different charges or helicities, and thus there are no spin-1 or spin-2 identical particles
which interact with themselves\,\cite{Nima}.}\, $h^3$ and $h^4$\,.\,
Among these, {\it the type-(ii) and type-(iii) are new forces
solely mediated by the Higgs boson itself.}\footnote{No other gauge bosons or fermions have
these features (cf.\ also footnote-1).
Such a spin-0 scalar Higgs boson holds a truly unique position
in the structure of the SM, and is far from being fully tested and understood.
In this sense, we can fairly regard, {\it the Higgs boson itself is new physics.}}\,
They are largely untested so far, and provide the most likely places to encode new physics
beyond the SM.
Even for the type-(i) force, the LHC Run-2 could only measure the $hWW$ and $hZZ$
couplings down to $(10-20)$\% at $2\sigma$ level \cite{Peskin}.
It should be stressed that {\it the discovery of the SM is not complete until all three types of
Higgs-involved forces are fully tested by direct measurements.}

The existence of such a spin-0 Higgs boson \,$h(125\text{GeV})$\, is truly profound.
This is because $\,h$\, is responsible for mass-generations for all SM particles,
the spin-1 weak gauge bosons and the spin-$\frac{1}{2}$ quarks and leptons,\footnote{%
The masses of active neutrinos can be naturally generated via seesaw mechanism
after including the right-handed singlet neutrinos, which still invokes Yukawa interactions
with the Higgs boson.}\,
via the above type-(i) and type-(ii) forces.
Note that the observed unnaturally large hierarchies among the quark and lepton masses
correspond to the same hierarchies among the Higgs Yukawa couplings.
These Yukawa couplings range from the top quark Yukawa coupling $\,y_t^{}\simeq 1\,$
down to a tiny electron Yukawa coupling $\,y_e^{}=\OO (10^{-6})$,\,
and have a rather irregular pattern,
which are all unexplained within the SM.
Hence, the Yukawa sector apparently calls for new physics.
The upper bounds on the new physics scales associated with all SM fermion mass-generations
vary within the range of $\,3.5-107$\,TeV, from the top quark to the electron \cite{Dicus:2004rg}.
This range of scales are mainly beyond the reach of the LHC, but are within the
(in)direct reaches of the next generation of high energy circular colliders
\cite{Nima}\cite{Dicus:2004rg}.
The Higgs boson $h$ also generates a physical mass
$\,M_h^{}\,$ for itself via its type-(iii) self-interaction force
after spontaneous symmetry breaking, but this mass is not protected
against radiative corrections, causing the naturalness problem \cite{HP}.
Furthermore, this Higgs boson could serve as the inflaton to drive
the required exponential expansion of the early universe \cite{HINF},
and may also be connected to dark matter \cite{DM}.
But, the SM Higgs potential suffers instability at scales well below the
Planck mass \cite{vacIn} and calls for new physics at or beyond the TeV scale \cite{TeV}.

The above physics considerations strongly motivate the next generation high energy colliders
beyond the LHC. Because of the profound implications of the newly discovered light Higgs boson
\,$h(125\text{GeV})$,\, it is natural to first precisely measure its properties at an
$e^+e^-$ Higgs factory and find compelling clues to the new physics.
There are three major proposals on the market,
the Circular Electron Positron Collider (CEPC) \cite{CEPC},
the Future Circular Collider (FCC-ee) \cite{FCC-ee},
and the International Linear Collider (ILC) \cite{ILC}.
All three proposed colliders can run at $\sqrt{s} = 250$\,GeV\, by producing Higgs boson via
Higgsstrahlung ($e^+ e^- \!\!\to Zh$) and
$WW$ fusion ($e^+ e^- \!\!\to \nu \bar \nu h$).
By measuring decay products of the final state $Z$ boson in the Higgsstrahlung process, the Higgs
signal can be extracted with the aid of recoil mass reconstruction technique \cite{recoil}. This allows
model-independent measurement of Higgs decay branching fractions down to percentage level.
The CEPC runs at the collision energy of 250\,GeV with 5\,ab$^{-1}$ integrated luminosity
can produce about 1 million Higgs bosons. With these, most Higgs decay channels can be
precisely measured with sizable events. Hence, such a Higgs factory will be an ideal place
to probe the new physics deviations via Higgs production and decays,
as well as other precision measurements.\footnote{Probing the Higgs self-interactions
is much harder at such a Higgs factory \cite{ee-h3}. But the Higgs self-coupling can be
measured via Higgs pair productions to good precision\,\cite{pp-h3} at the future
circular hadron colliders $pp$(100TeV) \cite{CEPC}\cite{FCC-ee}.
}

In this work, we will study the probe of new physics scales at the $e^+e^-$ Higgs
factory, with CEPC as a concrete example. For such a Higgs factory,
{\it all the new physics effects} associated with the light Higgs boson $h(125\text{GeV})$
can be parametrized by model-independent dimension-6 effective operators,
which involve the SM Higgs doublet $H$.
We will establish a scheme-independent approach
to optimize the constraints of new physics from all available observables,
including both the electroweak precision observables (EWPO) and
the Higgs observables (HO).  With this formulation, we treat the SM
electroweak parameters and the coefficients of dimension-6 operators
on equal footing for a combined analysis, which can be fitted simultaneously by
the same $\chi^2$ function. Since deviations from the SM are generally small,
we can expand the new physics parameters up to linear order and perform an
analytical $\,\chi^2\,$ fit to derive the potential reach of
the new physics scales. We will demonstrate that the $Z$-pole measurements
in the early phase of a Higgs factory can be as important as the Higgs observables.
Some aspects of the effects of these operators on
Higgsstrahlung production were studied before for $e^+e^-$ colliders at various
energies and with different focuses\,\cite{pre}, which usually did not cover a complete list
of these operators, and also did not consider the interplay
with precision observables. A recent paper\,\cite{Craig:2014una}
studied the probe of these operators at a Higgs factory in the $Z$-scheme by taking
the three most precisely measured electrowaek observables
(the fine structure constant $\alpha$, the Fermi constant $G_F^{}$, and the $Z$ boson
mass $M_Z^{}$) as fixed inputs. The extended studies considered the existing electroweak
precision observables at LEP \cite{Falkowski:2014tna},
and the measurements at a future Higgs factory \cite{Ellis:2015sca}.
But a comprehensive investigation to combine the constraints of all HO and EWPO
would be highly beneficial.

This paper is organized as follows. In \gsec{sec:2}, we first establish a scheme-independent
approach with linear expansion, which puts both dimension-6 operators and electroweak
parameters on equal footing for fitting the data.
In \gsec{sec:3}, we analyze the CP-conserving
dimension-6 effective operators that involve the SM Higgs doublet $H$,
and summarize their effects on the field redefinition, particle masses,
and interaction vertices, along with \gapp{sec:kinetic} on kinetic mixing of gauge bosons.
With these, in \gsec{sec:4}, we study the Higgs and precision observables
to deduce the reach of new physics scales that can be probed
at the $e^+e^-$ Higgs factory. Then, in \gsec{sec:5},
we present the reach of precision measurement of the SM Higgs couplings
at the CEPC, and apply this to study the probe of new physics scales associated with
the dimension-6 Yukawa-type operators.
Finally, we conclude in \gsec{sec:6}.
We also present our method of the analytic linear $\chi^2$ fit in \gapp{sec:chi2},
which is used for the current analyses in \gsec{sec:4} and \gsec{sec:5}.

\vspace*{3mm}
\section{Scheme-Independent Approach for Precision Observables}
\label{sec:2}
\vspace*{2mm}

The electroweak sector of SM contains three basic parameters,
the $SU(2)_L^{}$ gauge coupling $\,g$\,,\,
the $U(1)_Y^{}$ gauge coupling \,$g'$,\,
and the Higgs vacuum expectation value (VEV) $v$.
They can be determined by the existing precision tests,
especially the four most precisely measured observables: the $Z$ boson mass $\MZ$,\,
the $W$ boson mass $\MW$,\, the Fermi constant $\GF$,\,
and the fine structure constant $\alpha$\,.\,
Fixing the electroweak (EW) parameters $(g, g', v)$ needs only three observables as inputs.
In common practice, one usually adopts
either $Z$-scheme ($M_Z^{}, G_F^{}, \alpha$)
or $W$-scheme ($M_Z^{}, M_W^{}, \alpha$) to fix the values of $(g, g', v)$.\footnote{In
the literature\,\cite{Hollik86}, choosing the inputs ($M_Z^{}, G_F^{}, \alpha$) is also
called the intermediate scheme, and choosing the other set ($M_Z^{}, M_W^{}, \alpha$)
is called the on-shell scheme.}\,
Picking up which scheme is thus a matter of choice.
In addition, the numerical analysis with this approach
could only implement the central values of these electroweak observables
without including the associated uncertainties. This means, some information
from experimental measurements is discarded and the outcome turns out
to be scheme-dependent. In the present study,
we try to incorporate all precision observables, so we can realize
more sensitive probe of the new physics scales of the
dimension-6 operators by using all the information available from
experiments. The improvement of this new method is minor for the current precision data,
but will become significant for analyzing the future precision measurements
at the Higgs factory.

Our new strategy is to employ all the precision observables, including both their
central values and uncertainties. The values of \,$(g, g', v)$\, are determined by
fitting the data altogether with effective operator coefficients
[cf.~\geqn{eq:Lagrangian}], rather than
being expressed as functions of central values of 3 input parameters chosen for the $Z$-scheme
or $W$-scheme. In this way, we can utlize all the most precisely
measured precision EW observables ($M_Z^{},\, M_W^{},\, G_F^{},\, \alpha$) altogether
with any other relevant observables in our $\chi^2$ fit.
We no longer need to invoke the concept of scheme or input parameters.
Our analysis only involves model (fitting) parameters and experimental data.
The basic EW parameters and the coupling coefficients
can be treated equally as the fitting parameters.
This will just add three more fitting parameters,
but all precision observables can be equally used to constrain the dimension-6 operators.

An observable contains the SM contribution, expressed in terms of the
EW parameters, plus the corrections from new physics. When fitting experimental
measurements, the SM contribution and new physics contribution vary simultaneously.
So long as the precision measurements are included, the EW parameters are
constrained with small uncertainties. On the other hand, the new physics contribution
is expected to be small. Thus, it is well justified that both the EW parameters and
the effective operator coefficients have only small shifts from their reference values.
We can expand observables as linear combinations of the shifts.
In consequence, we can perform analytic $\chi^2$ fit as elaborated in \ref{sec:chi2}.

To implement the two features above, we treat all model-parameters (EW parameter and
effective operator coefficients) on equal footing and make analytic $\chi^2$ fit for small
variations. We lay out the procedure as follows. First, we split
each EW parameter \,$f$\, as a sum of the reference value
(acting as starting point for the $\chi^2$ fit)
and the shift from it,
\begin{equation}
  f^{(\text{sm})}
\,\equiv\,
  f^{(r)}
+ \delta f
\,\simeq\,
  f^{(r)}\!
\left(\!
  1
+ \frac {\delta f} f
\right) ,
\end{equation}
where $\,f^{(\text{sm})}\,$ is the SM prediction, $\,f^{(r)}\,$ the reference value, and
$\delta f$ the shift between them.
Note that before fitting the data each of these quantities exists only symbolically
and should be treated as a variable without specific value.
Then, any observable $\,X\,$ can be expanded in a similar way,
\begin{eqnarray}
X \,\equiv\,
  X \llbracket f^{(\text{sm})} \rrbracket
+ \overline{\delta X}
\,=\,
  X \llbracket f^{(r)} \rrbracket
+ X'\llbracket f \rrbracket \delta f
\!+ \overline{\delta X}
\,\equiv\,
  X^{(r)} \! + \widetilde{\delta X} ,
\end{eqnarray}
where $\,X \llbracket f \rrbracket$\,
is a functional form of the observable in terms of the EW
parameters, while the new physics contribution (which can arise
from the relevant dimension-6 operators \cite{EFT1, EFT2} for instance,
cf.\ Eq.\,\geqn{eq:Lagrangian} and Table\,\ref{tab:1}) and
the SM loop corrections are included in $\,\overline{\delta X}$\,.\, Corresponding
to the reference value $f^{(r)}$ of the EW parameters, the observable also
has a reference value
$\,X^{(r)} \!\equiv X \llbracket f^{(r)} \rrbracket$\,.\,
Its shift from reference value is then combined into
\beqa
\label{eq:dXt}
\widetilde{\delta X} \,\equiv\, X' \llbracket f \rrbracket \delta f +
\overline{\delta X} \,,
\eeqa
where $\,X' \llbracket f \rrbracket$\,
is a functional derivative with respect to $\,f\,$.

For the present study, we use $\,f = (\MZ,\,\GF,\, \alpha)$\,
as EW fitting parameters which are equivalent
to using $(g,\, g',\, v)$, but have the benefit of being direct physical observables.
Each of them contains a shift from its reference value,
\begin{equation}
M^{(\text{sm})}_Z \!=
M^{(r)}_Z \!\left(\! 1 \!+\! \frac {\delta M_Z}{M_Z} \right)\!,
~~~
G^{(\text{sm})}_F \!=
  G^{(r)}_F \!\left(\! 1 \!+\! \frac {\delta G_F}{G_F} \right)\!,
~~~
  \alpha^{(\text{sm})}\!
=
  \alpha^{(r)}\! \left(\! 1 \!+\! \frac {\delta \alpha} \alpha \right) \!.
\label{eq:Zscheme-inputs}
\end{equation}
The quantities \,$(\delta M_Z^{},\, \delta G_F^{},\, \delta\alpha)$\,
are the differences between the SM prediction and reference value.
Note that the SM prediction can be at either tree-level or loop-level, depending
on whether the loop-level correction needs to be taken into consideration.
In either case, the dependence on the electroweak parameter shifts remains the same,
since the SM loop-contribution is already at the linear order and hence can be treated
as a constant term in our linear $\chi^2$ fit,
\,$
\delta X_{\rm{1-loop}}^{} \llbracket f \rrbracket
\simeq
\delta X_{\rm{1-loop}}^{} \llbracket f^{(r)} \rrbracket
$\,,\,
as long as the reference values are close to the experimental central values
so that the effect of the shift $\,\delta f\,$ belongs to higher orders and
is negligible at the linear-order perturbation.
We will give an explicit example for the case of $W$ boson mass
in \gsec{sec:precision}, which justifies that Eq.\geqn{eq:Zscheme-inputs}
applies to either tree-level or loop-level analysis.\footnote{In the literature, the
choice of renormalization conditions on $(\alpha,\,\GF,\,\MZ)$ is also called
$Z$-scheme. Nevertheless, this $Z$-scheme is only for imposing renormalization conditions
and should not be confused with the $\chi^2$ fitting schemes,
especially the $Z$-scheme by fixing the values of $(\alpha,\,\GF,\,\MZ)$
as discussed in Sec.\,\ref{sec:2}.
Although $(\alpha,\,\GF,\,\MZ)$ are fixed to
their physical values by renormalization conditions,
the physical values themselves still have experimental uncertainties
and hence can be adjusted in $\chi^2$ fit. The ``fixing'' in a renormalization
condition only has symbolical meaning and does not remove the experimental uncertainty
of the corresponding physical value.}

In principle, the reference point \,$f^{(f)}$\, can take any value.
This arbitrariness is then compensated by the corresponding shift parameter
$\,\delta f$\,.\,  Nevertheless,
for our linear expansion and thus the analytic $\chi^2$ fit to work, the reference point should
be close to the best-fit value which is around the experimental central value.
When the reference value is fixed to the experimental central value and if no parameter is allowed
to be freely adjusted, the shift quantity $\,\widetilde{\delta X}$\, would vanish.
For our choice of $\,(\MZ,\, \GF,\, \alpha)$\, as fitting parameter,
the assumption of vanishing
\,$(\widetilde{\delta M_Z},\,\widetilde{\delta G_F},\,\widetilde{\delta \alpha})$\,
will reduce our scheme-independent approach to the commonly used $Z$-scheme.
For the practical analysis here,
we will assign the reference values of $\,(\MZ,\, \GF,\, \alpha)$\, to be their current
experimental central values for convenience and allow their shifts
$\,(\delta\MZ,\, \delta\GF,\, \delta \alpha)$\, to vary for scheme-independent fit
(when needed).

\vspace*{3mm}
\section{New Physics from Dimension-6 Effective Operators}
\label{sec:3}
\vspace*{2mm}

In this section, we first present the effective Lagrangian with relevant dimension-6 operators
for the current precision Higgs study (\gsec{sec:3.1}).
Then, we systematically derive their effects
via kinetic terms and mass terms (\gsec{sec:3.2}),
and via interaction vertices (\gsec{sec:3.3}).

\begin{table}[t]
\setlength{\tabcolsep}{0.7mm}
\centering
\caption{List of dimension-6 effective operators for the present study.}
\vspace*{3mm}
\begin{tabular}{ccc}
\hline\hline
  Higgs & EW Gauge Bosons & Fermions
\\
\hline
  $\mathcal O_H^{}\! = \frac 1 2 (\partial_\mu |H|^2)^2$
& $\mathcal O_{WW}^{}\! = g^2 |H|^2 W^a_{\mu \nu} W^{a\mu \nu}$
& $\mathcal O^{(3)}_L\! = (i H^\dagger\sigma^a\!\! \stackrel \leftrightarrow D_\mu^{}\!\! H)
(\overline \Psi_L^{} \gamma^\mu \sigma^a \Psi_L^{})$
\\[0.5mm]
  $\mathcal O_T^{}\! = \frac 1 2 (H^\dagger\! \stackrel\leftrightarrow D_\mu^{}\! H)^2$
& $\mathcal O_{BB}^{}\! = g^2 |H|^2 B_{\mu \nu} B^{\mu \nu}$
& $\mathcal O^{(3)}_{LL}\! =
  (\overline \Psi_L^{} \gamma_\mu \sigma^a \Psi_L^{})
  (\overline \Psi_L^{} \gamma^\mu \sigma^a \Psi_L^{})$
\\[0.5mm]
& $\mathcal O_{WB}^{}\! = g g' H^\dagger \sigma^a H W^a_{\mu \nu} B^{\mu \nu}$
& $\mathcal O_L^{}\! = (i H^\dagger\! \stackrel \leftrightarrow D_\mu^{}\! H)
(\overline \Psi_L^{} \gamma^\mu \Psi_L^{})$
\\[0.5mm]
  Gluon
& ~$\mathcal O_{HW}^{}\! = i g (D^\mu H)^\dagger \sigma^a (D^\nu H) W^a_{\mu \nu}$~
& $\mathcal O_R^{}\! = (i H^\dagger\!\! \stackrel \leftrightarrow D_\mu^{}\!\! H)
  (\overline \psi_R^{} \gamma^\mu \psi_R^{})$
\\ \cmidrule{1-1}
  $\mathcal O_g^{}\! = g_s^2 |H|^2 G^a_{\mu \nu} G^{a\mu \nu}$
& $\mathcal O_{HB}^{}\! = i g' (D^\mu H)^\dagger (D^\nu H) B_{\mu \nu}$
& $\mathcal O_y^u = |H|^2\,\overline{\Psi}_L^{q}\tilde{H}u_R^{}$
\\[0.5mm]
& $ \OO_{W}^{} =\fr{ig}{2}(H^\dagger\!\sigma^a\!\!\stackrel \leftrightarrow D_\mu^{}\!\! H)
                           D_\nu W^{a\mu\nu}$
& $\OO_y^d = |H|^2\,\overline{\Psi}_L^{q}{H}d_R^{}$
\\[0.5mm]
& $\OO_{B}^{} = \fr{ig'}{2}(H^\dagger\! \stackrel \leftrightarrow D_\mu^{}\!\! H)
                           D_\nu B^{\mu\nu}$
& $\OO_y^{\ell} = |H|^2\,\overline{\Psi}_L^{\ell}{H}\ell_R^{}$
\\
\hline\hline
\end{tabular}
\label{tab:O}
\label{tab:1}
\vspace*{3mm}
\end{table}

\vspace*{2mm}
\subsection{Effective Lagrangian with Dimension-6 Operators}
\label{sec:3.1}
\vspace*{2mm}

The new physics effects beyond the SM can be
generally parametrized by the dimension-6 effective operators
in a model-independent way \cite{EFT1,EFT2},
\begin{equation}
  \mathcal L
~=~
  \mathcal L_{\text{SM}}^{}
+ \sum_j \frac{c_j^{}}{\,\Lambda^2\,} \mathcal O_j^{} \,.
\label{eq:Lagrangian}
\end{equation}
If these new physics effects are associated with Higgs boson,
we expect a set of gauge-invariant and CP-conserving dimension-6 operators
will appear in the low energy effective theory, as summarized in Table\,\ref{tab:1}.
We expect the associated cutoff scale $\,\Lambda/{|c_j^{}|^{\frac{1}{2}}}\,$
to be around TeV scale or not far above it.
Since the physical processes at an $e^+e^-$ Higgs factory with
$\,\sqrt{s}=240-250$\,GeV\, have energy scales well below the TeV scale,
we see that the effective Lagrangian \eqref{eq:Lagrangian} provides a
perfectly valid low energy formulation of the new physics effects.

This effective Lagrangian contains ten bosonic and seven fermionic dimension-6 operators,
where each operator is associated with its own coefficient $\,c_j^{} / \Lambda^2$.\,
We note that integration by part gives the identities \cite{eom},
\begin{subequations}
\label{eq:ID}
\begin{eqnarray}
\label{eq:ID-B}
  \mathcal O_B^{}
& = &\,
  \mathcal O_{HB}^{}
+ \frac 1 4 \(\mathcal O_{BB}^{} \!+ \mathcal O_{WB}^{}\) ,
\\[1mm]
\label{eq:ID-W}
  \mathcal O_W^{}
& = &
  \mathcal O_{HW}^{}
+ \frac 1 4 \(\mathcal O_{WW}^{} \!+ \mathcal O_{WB}^{}\) .
\end{eqnarray}
\end{subequations}
It means that among the seven operators
($\mathcal O_B^{}$, $\mathcal O_W^{}$, $\mathcal O_{BB}^{}$,
 $\mathcal O_{WB}^{}$, $\mathcal O_{WW}^{}$,
 $\mathcal O_{HB}^{}$, $\mathcal O_{HW}^{}$), two of them are redundant.
We could use these to eliminate ($\OO_B^{},\,\OO_W^{}$), which is called the HISZ basis
in the literature \cite{basis}.
We further note that the operators ($\OO_B^{},\,\OO_W^{}$) can be replaced by
the equation of motion (EOM) \cite{eom},
\beqs
\label{eq:EOM}
\beqa
\label{eq:EOM-B}
\mathcal O_B^{} & = &\, {{g'}^2}\!\left[
-\fr{1}{2}\OO_T^{} + \fr{1}{2}(Y_L^{}\OO_L^{} +Y_R^{}\OO_R^{})
\right] \!,
\\[1mm]
\label{eq:EOM-W}
\mathcal O_W^{} & = &\, g^2\!\left[
-\fr{3}{2}\OO_H^{} + 2\OO_6^{} + \fr{1}{2}\(
\OO_y^u +\OO_u^d +\OO_y^{\ell} +\textrm{h.c.}\)
+\fr{1}{4}\OO_L^{(3)} \right]\!,
\eeqa
\eeqs
where $\,Y_L^{}\,$ and $\,Y_R^{}\,$ stand for the hypercharges of fermion fields
$\Psi_L^{}$ and $\Psi_R^{}$, respectively.
Eqs.\,\eqref{eq:EOM-B}-\eqref{eq:EOM-W} also make ($\OO_B^{},\,\OO_W^{}$) redundant.
Thus, one may use the identities \eqref{eq:ID-B}-\eqref{eq:ID-W} to eliminate the
two other operators ($\OO_{WW}^{},\,\OO_{WB}^{}$) instead. This means that the four operators
($\OO_B^{},\,\OO_W^{}$) and ($\OO_{WW}^{},\,\OO_{WB}^{}$) are redundant,
and can be eliminated in principle.
For the current first-step study, with the limited experimental observables and a large number
of dimension-6 effective operators, we will not carry out a global $\chi^2$ fit of all operators
together. Instead, we perform the $\chi^2$ fit by {including only one operator at each time,}
which is common in the literature.
So we need not to exclude the redundant operators. In this way, we can first examine how each operator
contributes and how it can be constrained, for completeness.
Nevertheless, in the current study,
we will always impose the basic identities \eqref{eq:ID-B}-\eqref{eq:ID-W} to eliminate
($\OO_B^{},\,\OO_W^{}$), as is the commonly used HISZ basis\,\cite{basis}.
[But we stress that when considering a future global $\chi^2$ fit including many operators simultaneously,
 it is necessary to remove all the redundant operators by using both the identities
 \eqref{eq:ID} and the EOM \eqref{eq:EOM}.]

Note that in Table\,\ref{tab:1},
$\mathcal O^{(3)}_{LL}$ does not involve the SM Higgs doublet, but we
take it into account since it affects the Fermi constant (which is
the coefficient of dimension-6 four-fermion operator itself),
and consequently the other observables through parameter shift.
Since each of the Yukawa-type effective operators $(\OO_y^u,\,\OO_y^d,\,\OO_y^{\ell})$
modifies the SM Yuakawa coupling only by a rescaling factor, we study their tests
separately in \Ssec{sec:5}.

If the underlying UV theory for these effective operators is known, their coefficients
could be expressed in terms of the model-parameters in principle.
For the present study, we follow the model-independent effective theory formulation,
where the coefficients of dimension-6 operators are independent of each other.
We will use experimental measurements to estimate the potential reach of
indirectly probing the effective new physics scale
$\,\Lambda_j^{} \equiv \Lambda/{|c_j^{}|^{\frac{1}{2}}}\,$
associated with each operator at the Higgs factory
(cf.\ \gsec{sec:4}).\footnote{Some recent studies of specific new physics models
or specific processes at Higgs factories appeared in Ref.\,\cite{otherx}.}\,
We will keep in mind that the coefficient $\,c_j^{}\,$ of each effective operator
$\,\OO_j^{}\,$ usually depends on powers of the couplings from the underlying UV theory,
which could be larger or smaller than $\,\OO(1)$.\, The coefficient $\,c_j^{}\,$
could also depend on loop-factors when $\,\OO_j^{}\,$ is induced from loop-level
contributions (such as the case of $\,\OO_g^{}\,$). Hence, it is a model-dependent issue
to further convert our general bound on $\,\Lambda_j^{}\,$ to the corresponding bound on
$\,\Lambda\,$.\,
In the rest of this section, we first analyze the contributions of these
dimension-6 operators to the relevant Feynman vertices and physical observables.

\vspace*{2mm}
\subsection{New Physics via Kinetic Terms and Mass Terms}
\vspace*{1mm}
\label{sec:3.2}

Before drawing Feynman diagrams and computing the relevant Higgs production cross sections
and decay width, it is necessary to check whether all the involved propagators take
their canonical form.  If not, we need to make proper field redefinitions as summarized in
\gapp{sec:kinetic}.
Such redefinitions will modify the relevant mass terms and interaction vertices.
With the dimension-6 operators in \gtab{tab:O}, the kinetic terms of fermions
remain the same, while those of bosonic fields, the Higgs field $h$ and the gauge bosons
($W^\pm,\, Z^0,\, A^0$), are affected.

\vspace*{1mm}
\subsubsection{Higgs Field $h$}
\vspace*{1mm}

The operator $\mathcal O_H$ in \gtab{tab:O} could contribute a nonzero correction
to the kinetic term of the Higgs field. Redefining the Higgs field,
\begin{equation}
  h
~\rightarrow~
\left(
  1
- \frac 1 2 \frac {v^2}{\Lambda^2} c_H
\right) h \equiv   Z_h h ,
\label{eq:h-redef}
\end{equation}
will absorb the deviation from the canonical form.
It applies to every $h$ that appears in the Lagrangian and leads to a rescaling factor
for any interaction vertex involving Higgs field(s). Each Higgs field $h$ receives a
rescaling factor $Z_h$, and Higgs mass term receives a rescaling factor $Z^2_h$.

\vspace*{1mm}
\subsubsection{Charged Gauge Boson}
\vspace*{1mm}

The $W^\pm$ gauge bosons receive a correction to its kinetic term from
the operator $\mathcal O_{WW}$ in \gtab{tab:O}.
This leads to the field redefinition of the $W^\pm$ bosons,
\begin{equation}
  W^\pm ~\to~
\left(
  1 +  \frac {v^2}{\Lambda^2} g^2 c_{WW}^{}
\right) W^\pm
\,\equiv\,
  Z_W W^\pm .
\label{eq:W-redef}
\end{equation}
Although the $W$ mass receives no direct correction, the field redefinition and parameter shift
can contribute indirectly,
\begin{equation}
  \frac {\widetilde{\delta \MW}}{\MW}
\,=\,
  \frac 1 {c^2_w - s^2_w}
\left[
  c^2_w \frac {\delta \MZ}{\MZ}
+ \frac 1 2 s^2_w
  \left(
    \frac {\delta \GF}{\GF}
  - \frac {\delta \alpha} \alpha
  \right)
\right]
+ \frac {v^2}{\Lambda^2} g^2 c_{WW}^{} \,,
\label{eq:W-del-mass}
\end{equation}
according to \geqn{eq:w-redef-m2}. The weak mixing angle is denoted as
$(c_w, s_w) \equiv (\cos \theta_w, \sin \theta_w)$ evaluated at the reference point.
Note that the correction from field redefinition to the mass term has the same sign
as in \geqn{eq:W-redef}.

\vspace*{1mm}
\subsubsection{Neutral Gauge Bosons}
\vspace*{1mm}

The case of neutral gauge bosons is a little bit more complicated since both kinetic
term and mass term are $2 \times 2$ matrices.
From the dimension-6 operators
$(\mathcal O_{WW}, \mathcal O_{BB}, \mathcal O_{WB})$
in \gtab{tab:O}, we derive corrections to the kinetic term,
$\mathbb I \partial^2 \rightarrow \mathbb K \partial^2 \equiv (\mathbb I + \delta K)$,
with the explicit form of $\delta K$ given in \geqn{eq:dK}.
The neutral $Z$ and $A$ gauge bosons need to be not only redefined but also diagonalized
as we elaborate in \gapp{sec:kinetic-NC},
\begin{subequations}
\begin{eqnarray}
  Z^\mu
& \rightarrow &
\left[
  1
+  \frac {v^2}{\Lambda^2}
\left(
 c^2_w g^2 c_{WW}^{} + c_w^{} s_w^{} g g' c_{WB}^{} + s^2_w g'^2 c_{BB}^{}
\right)
\right]\! Z^\mu ,
\label{eq:Z-redef}
\\
  A^\mu
& \rightarrow &
\left[
  1
+  \frac {v^2}{\Lambda^2}
\left(
s^2_w g^2 c_{WW}^{} - c_w^{} s_w^{} g g' c_{WB}^{} + c^2_w g'^2 c_{BB}^{}
\right)
\right]\! A^\mu
\nonumber
\\
&& +
  2\frac {v^2}{\Lambda^2}
  \left[ c_w^{} s_w^{} g^2 c_{WW}^{} - \frac 1 2 (c^2_w - s^2_w) g g' c_{WB}^{}
  - c_w^{} s_w^{} g'^2 c_{BB}^{} \right]\! Z^\mu .
\label{eq:A-redef}
\end{eqnarray}
\label{eq:NC-redef}
\end{subequations}
For convenience, we denote the field redefinition of $A$ and $Z$ as
$A \rightarrow Z_A^{} A + \delta Z_X^{} Z \equiv (1 + \delta Z_A^{}) A + \delta Z_X^{} Z$
and
\,$Z \rightarrow Z_Z^{} Z \equiv (1 + \delta Z_Z^{}) Z$,\, respectively,
where the explicit form of
\,$(\delta Z_A^{},\, \delta Z_Z^{},\, \delta Z_X^{})$\,
can be read off from \geqn{eq:NC-redef}.
Note that kinetic mixing can introduce not only field redefinition $(\delta Z_A, \delta Z_Z)$ to both
$A$ and $Z$, but also equal correction $\delta Z_X$ to the left- and right-handed currents
of the $Z$ boson from the electromagnetic current as shown in the last line.

Any vertex involving $n$ fields of $Z$ should be divided by a factor of
$Z^n_Z$ due to this field redefinition. The mass term can be treated as
a vertex with two gauge fields. Hence, it is rescaled by $Z^2_Z$.
The mass of the neutral gauge boson $Z$ is also affected by $\mathcal O_T$
in \gtab{tab:O},
\begin{equation}
  \frac {\widetilde{\delta \MZ}}{\MZ}
=
  \frac {\delta \MZ}{\MZ}
- \frac 1 2 \frac {v^2}{\Lambda^2} c_T^{}
+ \delta Z_Z^{} \,.
\label{eq:Z-del-mass}
\end{equation}
where the extra contribution comes from the field redefinition \geqn{eq:Z-redef} of
the $Z$ gauge boson as indicated by the general analysis in \gapp{sec:kinetic-NC}.

\vspace*{1mm}
\subsubsection{Gluons}

Once the Higgs field $H$ develops nonzero VEV, the operator $\mathcal O_g$ in
\gtab{tab:O} can induce a correction to the kinetic term of gluons. The effect
is a field redefinition,
\begin{equation}
  G^a_\mu
\rightarrow
\left(
  1
+  \frac {v^2}{\Lambda^2} g^2_s c_g
\right) G^a_\mu
\equiv
  Z_G G^a_\mu \,,
\label{eq:g-redef}
\end{equation}
which only affect the relevant interaction vertices.

\vspace*{3mm}
\subsection{New Physics via Interaction Vertices}
\vspace*{1.5mm}
\label{sec:3.3}

The new physics parameters of the dimension-6 operators can affect
the interaction vertices in three ways.
First, they can give direct contributions to the existing vertex,
sometimes with a different tensor structure such as the case of $ZZh$ coupling.
Second, the field redefinition can introduce an overall rescaling factor of
the relevant vertex that contains the corresponding field.
Finally, the shifts of electroweak parameters from their reference values can affect
the existing vertex through zeroth order correlations.
In addition, the dimension-6 operators may introduce some new vertices, such as
the trilinear vertex $AZh$, and other quartic interactions $\,Zh \psi \bar \psi\,$
and $\,W h \ell \nu$.

\vspace*{1mm}
\subsubsection{Gauge Boson Coupling with Fermions}
\label{sec:Zff}
\label{sec:Wff}
\vspace*{1mm}

The coupling between the charged gauge boson $W^\pm$ and leptons can be modified by
the operator $\mathcal O^{(3)}_L$ in \gtab{tab:O} and
the $W$ field redefinition \geqn{eq:W-redef},
\begin{equation}
  \frac g {\sqrt 2}
\left(
  \frac {v^2}{\Lambda^2} c^{(3)}_L
+ \frac {v^2}{\Lambda^2} g^2 c_{WW}^{}
\right)
  \(W^+_\mu \bar \nu_L^{} \gamma^\mu \ell_L^{}
    + W^-_\mu \bar \ell_L^{} \gamma^\mu \nu_L^{} \) .
\end{equation}
Note that the direct correction to this vertex has the same form as the SM counterpart.
Hence, its contribution can be combined into the overall coupling constant.
In addition, $\,g_{W\ell \nu}^{}\,$ can split into the reference value plus
parameter shift in $\,g\,$,
\begin{equation}
  \frac {\widetilde{\delta g_{W \ell \nu}^{}}}{g_{W \ell \nu}^{}}
~=~
\frac{1}{\,\cos 2\theta_w^{}\,}
\left(
  c^2_w \frac {\delta \MZ}{\MZ}
+ \frac 1 2 c^2_w \frac {\delta \GF}{\GF}
- \frac 1 2 s^2_w \frac {\delta \alpha} \alpha
\right)
+ \frac {v^2}{\Lambda^2} c^{(3)}_L
+ \frac {v^2}{\Lambda^2} g^2 c_{WW}^{} .
\label{eq:Wenu-del}
\end{equation}

For the $Z \bar \ell \ell$ vertex, new physics contributions arise from both direct correction
and kinetic mixing. The first part comes from operators
$(\mathcal O^{(3)}_L, \mathcal O_L, \mathcal O_R)$ in
\gtab{tab:O},
\begin{equation}
  \frac g {2 c_w} \frac {v^2}{\Lambda^2}
\left[
  \left( c^{(3)}_L \!- c_L \right) \bar \nu_L \gamma^\mu \nu_L
- \left( c^{(3)}_L \!+ c_L \right) \bar \ell_L \gamma^\mu \ell_L
- c^\nu_R \bar \nu_R \gamma^\mu \nu_R
- c^\ell_R \bar \ell_R \gamma^\mu \ell_R
\right] Z_\mu^{} .
\label{eq:zff}
\end{equation}
We can see that the four terms are independent of each other with four different
dimension-6 operator coefficients.
In addition, the redefinitions \geqn{eq:NC-redef} of $(Z,A)$
introduce extra corrections to the left- and right-handed currents,
\begin{subequations}
\begin{eqnarray}
  \delta g^*_L & = &
  Q\, g_z^{} c_w^{} s_w^{} \delta Z_X^{}
+ g_z^{} (T_3^{} - s^2_w Q) \delta Z_Z^{} ,
\\[1.5mm]
  \delta g^*_R & = &
  Q\, g_z^{} c_w^{} s_w^{} \delta Z_X^{}
- g_z^{} s^2_w Q \delta Z_Z^{} .
\end{eqnarray}
\end{subequations}
The first term is universal for left- and right-handed couplings, since
it comes from the $Z$-$A$ mixing and most importantly is proportional to
the electromagnetic current. On the other hand, the
second term comes from the field redefinition of the $Z$ gauge boson,
rendering it proportional to the SM prediction of $g_L^{}$ and $g_R^{}$, respectively.
Finally, from the zeroth-order coupling, extra correction can appear through parameter shift.
Here we show the correction to the coupling with charged leptons,
\begin{subequations}
\begin{eqnarray}
  \widetilde{\delta g_L}
& \equiv &
- \left[
  \frac 1 {2 \cos 2 \theta_w}
\left(
  \frac {\delta \MZ}{\MZ}
+ \frac 1 2 \frac {\delta \GF}{\GF}
\right)
- \frac {c^2_w s^2_w}{\cos 2 \theta_w^{}}
  \frac {\delta \alpha} \alpha
\right]\! g_z^{}
- \frac{g_z^{} v^2}{2 \Lambda^2} \left( c^{(3)}_L + c_L^{} \right)
+ \delta g^*_L ,
\qquad
\quad
\\
  \widetilde{\delta g_R^{}}
& \equiv &
- \left[
  \frac {s^2_w} {\cos 2 \theta_w}
\left(
  \frac {\delta \MZ}{\MZ}
+ \frac 1 2 \frac {\delta \GF}{\GF}
\right)
-  \frac {c^2_w s^2_w}{\cos 2 \theta_w^{}}
  \frac {\delta \alpha} \alpha
\right]\! g_z^{}
- \frac{g_z^{} v^2 }{2 \Lambda^2} c_R^{}
+ \delta g^*_R ,
\end{eqnarray}
\end{subequations}
where the second term accounts for the direct contribution summarized in \geqn{eq:zff}.
For convenience, we use $\,g_z^{} \equiv g/\cos \theta_w^{}\,$
to denote the weak gauge coupling associated with $Z$ boson.

\vspace*{1mm}
\subsubsection{Gauge Boson Couplings with Higgs}
\label{sec:ZZh}
\label{sec:WWh}
\label{sec:AZh}
\label{sec:hAA}
\label{sec:hgg}
\vspace*{1mm}

Corrections to the $ZZh$ vertex arise from
$(\mathcal O_T^{}, \mathcal O_{WW}^{}, \mathcal O_{BB}^{}, \mathcal O_{WB}^{},
  \mathcal O_{HW}^{}, \mathcal O_{HB}^{})$ in \gtab{tab:O},
\begin{equation}
- \frac {g^2 v}{2 c^2_w} \frac {v^2}{\Lambda^2} c_T^{} h Z_\mu^{} Z^\mu
\hspace{-1mm}
+
  \delta Z_Z^{} h \mathcal Z_{\mu \nu}^{} \mathcal Z^{\mu \nu}
+
  \frac g 2 \frac {v \partial_\mu h}{\Lambda^2}
\left[
  g c_{HW}^{}
+ \frac {s_w^{}}{c_w^{}} g' c_{HB}^{}
\right] Z_\nu^{} \mathcal Z^{\mu \nu}
,~~~
\label{eq:ZZh-1}
\end{equation}
where $\mathcal Z_{\mu \nu}^{} \equiv \partial_\mu^{} Z^{}_\nu - \partial_\nu^{} Z_\mu^{}$.\,
Note that the Higgs field redefinition \geqn{eq:h-redef} also contributes an overall term,
$- \delta Z_h^{} g_z^{} M_Z^{} \frac 1 2 h Z_\mu^{} Z^\mu $,\,
which should be combined with the first term that has the same tensor structure as
the SM contribution. To keep the expression neat, let us define
$\,f^{\mu \nu} \llbracket p,q \rrbracket \equiv p^\nu q^\mu - (p \cdot q) g^{\mu \nu}$.\,
Then, the Feynman rule reads,
\begin{eqnarray}
  i g_{ZZh}^{}
\left\{
  \left( 1 + c^Z_0 \right) g^{\mu \nu}
+ c^Z_1 f^{\mu \nu} \llbracket k_1^{}, k_2^{} \rrbracket
+ c^Z_2 f^{\mu \nu} \llbracket k_1^{}, k_1^{} \rrbracket
+ c^Z_3 f^{\mu \nu} \llbracket k_2^{}, k_2^{} \rrbracket
\right\} ,
\quad
\label{eq:fr-ZZh-2}
\end{eqnarray}
with $\,g_{ZZh}^{} \equiv g \MZ / c_w^{}$.\,
The decomposition \geqn{eq:fr-ZZh-2} is useful when discussing Higgs decay and will
be applied to the $W^+ W^- h$ and $AZh$ vertices discussed later in this section.
The coefficients $c^Z_i$ are defined as
\begin{subequations}
\begin{eqnarray}
c^Z_1 & = &  \frac 2 {\Lambda^2}
\left[
  4 \left( s^4_w c_{BB}^{} + c^2_w s^2_w c_{WB}^{} + c_w^4 c_{WW}^{} \right)
- \left( s^2_w c_{HB}^{} + c^2_{HW} \right)
\right] ,
\\
  c^Z_2 & = &  c^Z_3 =
- \frac 1 {\Lambda^2} \left( s^2_w c_{HB}^{} + c^2_w c_{HW}^{} \right) .
\end{eqnarray}
\label{eq:fr-ZZh}
\end{subequations}
Note that the overall rescaling factor of the SM contribution has been combined
with the $Z$ boson field redefinition \geqn{eq:Z-redef} and the parameter shift,
\begin{eqnarray}
c^Z_0
\,=\,  \frac {\widetilde{\delta g_{ZZh}^{}}}{g_{ZZh}^{}}
\,=\,  \frac 1 2 \frac {\delta \GF}{\GF}
+ 2 \frac {\delta \MZ}{\MZ}
- 2 \frac {v^2}{\Lambda^2} \left( c_T^{} + \frac 1 4 c_H^{} \right)
+ 2 \delta Z_Z^{} \,.
\label{eq:del-gZZh}
\end{eqnarray}

The $W^+ W^- h$ vertex is much simpler without complication from kinetic mixing.
It receives corrections from $(\mathcal O_{WW}^{},\, \mathcal O_{HW}^{})$
in \gtab{tab:O},
\begin{equation}
  2 g^2 \frac {v h}{\Lambda^2} c_{WW}^{} \mathcal W^+_{\mu \nu} \mathcal W^{-\mu \nu}
- \frac {g^2} 2 \frac {v \partial_\mu^{} h}{\Lambda^2} c_{HW}^{}
  ( W^+_\nu \mathcal W^{- \mu \nu} + W^-_\nu \mathcal W^{+ \mu \nu}) ,
\end{equation}
where $\mathcal W^{\pm \mu \nu} \equiv \dif^\mu W^{\pm\nu}-\dif^\nu W^{\pm\mu}$.
It can be grouped into the same form as \geqn{eq:fr-ZZh-2}, with $p_\pm$
denoting the momenta of $W_\pm^{}$,
\begin{equation}
  i g M_W
\left\{
  \left( 1 + c^W_0 \right)
  g^{\mu \nu}
+ c^W_1 f^{\mu \nu} \llbracket p_+^{}, p_-^{} \rrbracket
+ c^W_2 f^{\mu \nu} \llbracket p_+^{}, p_+^{} \rrbracket
+ c^W_3 f^{\mu \nu} \llbracket p_-^{}, p_-^{} \rrbracket
\right\} ,
\label{eq:FR-hWW-1}
\end{equation}
where the coefficients $c^W_i$ are defined as
\begin{equation}
  c^W_1
=  \frac 2 {\Lambda^2} \left( 4 c_{WW}^{} \!+ c_{HW}^{} \right) ,
\quad~~~
  c^W_2 = c^W_3
=
  \frac 1 {\Lambda^2} c_{HW}^{} .
\label{eq:cWs}
\end{equation}
The field redefinitions of $W$ and Higgs field
redefinitions, \geqn{eq:W-redef} and \geqn{eq:h-redef},
contribute as an overall rescaling and hence
can be combined with the parameter shift,
\begin{eqnarray}
  c^W_0
\,=\,
  \frac {\widetilde{\delta g_{WWh}^{}}}{g_{WWh}^{}}
\,=\,
  \frac 1 2
\left(
  2 c^2_w \frac {\delta \MZ}{\MZ}
+ \frac 1 2 \frac {\delta \GF}{\GF}
- s^2_w \frac {\delta \alpha} \alpha
\right)
- \frac 1 2 \frac {v^2}{\Lambda^2} c_H^{}
+ 2 \frac {v^2}{\Lambda^2} g^2 c_{WW}^{} .~~~~
\label{eq:gWWh-del}
\end{eqnarray}

In the SM, the photon $A_\mu^{}$ only couples
to a pair of charged particle and its anti-particle.
This is violated by effective operators
$(\mathcal O_{WW}^{}, \mathcal O_{BB}^{}, \mathcal O_{WB}^{}, \mathcal O_{HW}^{},
  \mathcal O_{HB}^{})$ in \gtab{tab:O},
\begin{equation}
2 \frac {\delta Z_X^{}} v h\mathcal Z_{\mu \nu}^{} F^{\mu \nu}
+ \frac {s_wg^2v}{2c_w\Lambda^2}
\left(
  c_{HW}^{}
- c_{HB}^{}
\right) \partial_\mu^{} h Z_\nu^{} F^{\mu \nu} ,
\end{equation}
where $\,F^{\mu\nu}=\dif^\mu A^\nu -\dif^\nu A^\mu$\,
is the field strength of photon.
We can see that the first term actually comes from kinetic mixing which is proportional to
$\,\delta Z_X^{}\,$ and arises from the second line of \geqn{eq:A-redef}.
With everything combined, the Feynman rule of this vertex
$A^\mu Z^\nu h$ can be grouped into,
\begin{subequations}
\begin{equation}
  i g_{ZZh}^{}
\left(
  c^A_1 f^{\mu \nu} \llbracket k_A^{}, k_Z^{} \rrbracket
+ c^A_3 f^{\mu \nu} \llbracket k_A^{}, k_A^{} \rrbracket
\right) ,
\end{equation}
where
\begin{equation}
  c^A_1
+ c^A_3
=
  2 \frac {\delta Z_X^{}} {\,g_{ZZh}^{} v\,} ,
\qquad
  c^A_3
= - \frac v {\,\Lambda^2\,} c_w^{} s_w^{} (c_{HW}^{} - c_{HB}^{}) .
\end{equation}
\label{eq:fr-AZh}
\end{subequations}

In SM, the Higgs boson couples with a pair photons/gluons through triangle
loops. The $hAA$ and $hgg$ vertices can also be induced
from high-energy theory, and can be contributed by the effective dimension-6
operators. From the operators $\mathcal O_{WW}^{}$, $\mathcal O_{BB}^{}$,
and $\mathcal O_{WB}^{}$ in \gtab{tab:O}, the Higgs field $h$ can
directly couple with a pair of photons with the effective coupling,
\beqa
V_{hAA}^{}
\,=~ {\frac{4}{\,v\,}\,\delta Z_A^{}}\,
  f^{\mu \nu} \llbracket p_1^{}, p_2^{} \rrbracket ,
\label{eq:FR-hAA}
\eeqa
where the momenta are assigned as $\,A^\mu(p_1^{}) A^\nu(p_2^{})\,h\,$.\,
The operator $\,\mathcal O_g^{}$\, induces the effective coupling,
\beqa
  V_{hgg}^{} \,=~ \frac {4v}{\,\Lambda^2\,} g_s^2 c_g^{} \delta_{ab}^{}
  f^{\mu \nu} \llbracket p_1^{}, p_2^{} \rrbracket ,
\label{eq:FR-hgg}
\eeqa
for the vertex $\,g^\mu(p_1^{}) g^\nu(p_2^{}) h$\,.\,
Note that the above tree-level corrections by the dimension-6 operators should be of
the same order as the one-loop contributions in the SM.

\vspace*{2mm}
\subsubsection{Hybrid Couplings between Bosons and Fermions}
\label{sec:Zhff}
\label{sec:Whff}
\vspace*{1mm}

The first vertex $\,Z h \bar f f\,$ arises from
$\,(\mathcal O^{(3)}_L, \mathcal O_L^{}, \mathcal O_R^{})$\,
in \gtab{tab:O},
\begin{equation}
  \frac{\,g_z^{}v\,}{\,\Lambda^2\,}
\left[
  \left( c^{(3)}_L \!- c_L^{} \right) Z_\mu^{} \bar u_L^{} \gamma^\mu u_L^{}
- \left( c^{(3)}_L \!+ c_L^{} \right) Z_\mu^{} \bar d_L^{} \gamma^\mu d_L^{}
- c^\psi_R \bar \psi_R \gamma^\mu \psi_R
\right]\! h \,.
\end{equation}
The corresponding Feynman rules are
\begin{subequations}
\begin{eqnarray}
  \bar u u Z h\!:
&&
 i \frac {g_zv}{\Lambda^2} \gamma^\mu
\left[
+ \left( c^{(3)}_L \!- c_L^{} \right) P_L^{}
- c^u_R P_R^{}
\right] ,
\\
  \bar d d Z h \!:
&&
 i \frac{g_z^{}v}{\Lambda^2} \gamma^\mu
\left[
- \left( c^{(3)}_L \!+ c_L^{} \right) P_L^{} - c^d_R P_R^{}
\right] .
\end{eqnarray}
\label{eq:qqZh}
\end{subequations}
Similarly, the vertex $\,W^+ h \bar f f'\,$
can arise from $\mathcal O^{(3)}_L$,
\begin{equation}
  \frac {\sqrt 2 g v c^{(3)}_L}{{\Lambda^2}}
\left(
  W^+_\mu \bar u_L^{} \gamma^\mu d_L^{}
+ W^-_\mu \bar d_L^{} \gamma^\mu u_L^{}
\right) h \,.
\label{eq:Wenuh}
\end{equation}

\vspace*{3mm}
\section{Probing New Physics Scales of Dimension-6 Operators}
\label{sec:4}
\vspace*{2mm}

The dimension-6 operators in \gtab{tab:O} can contribute to a wide range
of physical observables, including the electroweak precision observables (EWPO)
and the Higgs observables (HO) at a Higgs factory. Using the scheme-independent
approach, we can utilize all of them to constrain the dimension-6 operators.
Both the EWPO and HO could sensitively probe the
new physics at high energy \cite{Ellis-LHC-TGC,Henning:2014gca,Falkowski:2014tna,Ellis:2015sca}.
For instance, Ref.\,\cite{Ellis-LHC-TGC} studied the LHC Run-1 constraints on some dimension-6 operators
via measurements of triple gauge couplings, while Ref.\,\cite{Falkowski:2014tna}
studied the LEP-I and LEP-II limits on the coefficients of dimension-6 operators.
These can probe the new physics scales of dimension-6 operators from roughly a TeV up to
about 10\,TeV.

In \gsec{sec:precision} and \gsec{sec:collider}, we first derive the
contributions of dimension-6 operators to precision observables
$(\alpha, \GF, \MZ, \MW)$ and Higgs observables (among which two production cross sections
$\sigma(Zh)$ and $\sigma(h \nu \bar \nu)$ together with all decay branching fractions
can be measured).
Then, we use these results, supplemented by the existing precision measurements,
to estimate the new physics scales that can be probed at the CEPC in \gsec{sec:scale}.
We show the CEPC probe of these new physics scales can reach up to 10\,TeV.
We continue to elaborate the role of precision observables in
\gsec{sec:EWPO}, and demonstrate that the much more precisely measured $(\alpha, \GF, \MZ)$
effectively fix the three EW parameters $(g,\, g',\, v)$,\, while the less precisely
known $\,\MW\,$ helps to enhance the new physics scale limit. The situation changes
if $\,\MW\,$ can achieve comparable precision with $\,\MZ\,$ at Higgs factory as
demonstrated in \gsec{sec:ZWmass}. We include more precision observables at
$Z$-pole running of the $e^+e^-$ Higgs factory in \gsec{sec:Zpole},
and demonstrate that the limit on the new physics scale
can be further pushed up to around $30\,\mbox{TeV}$.

\vspace*{2mm}
\subsection{New Physics Contributions to Precision Observables}
\label{sec:precision}
\vspace*{1.5mm}

The existing best electroweak measurements include the weak gauge boson masses
$(M_W^{},M_Z^{})$, the fine-structure constant $\alpha$,
and the Fermi constant $\,\GF$\,.\, Since they have already been measured
experimentally, it is necessary to consider both their central values and uncertainties.
To achieve this, we will include the SM loop-corrections (which are of the same order as the dimension-6
operators) altogether. In this subsection, we first show how
the four precision observables $(\alpha, \GF, \MZ, \MW)$ are affected by dimension-6
operators via their linear combination and by the SM one-loop corrections via a constant term.
Since we have four observables versus three electroweak parameters, only one observable ($\MW$)
will receive explicit SM loop correction if the other three $(\alpha,\, \GF,\, \MZ)$\, are used to
fix the renormalization conditions.

\vspace*{2mm}
\subsubsection{Fine-Structure Constant}
\vspace*{1mm}

The fine-structure constant rescales by the photon field redefinition \geqn{eq:A-redef},
$\overline{\delta \alpha}/\alpha = 2 \delta Z_A^{}$.\, In addition,
the parameter shift can also induce a correction. Altogether we have,
\begin{equation}
  \frac {\widetilde{\delta \alpha}} \alpha
\,\simeq\,
  \frac {\delta \alpha} \alpha
+ 0.0111
\left(
  \frac {c_{WW}^{}}{\LambdaTeV^2}
- \frac {c_{WB}^{}}{\LambdaTeV^2}
+ \frac {c_{BB}^{}}{\LambdaTeV^2}
\right) ,
\label{eq:alpha}
\end{equation}
where $\,\LambdaTeV \equiv \Lambda/\mbox{TeV}\,$ is the cutoff scale in unit of TeV.
Since the measurement of the fine-structure constant $\,\alpha\,$ is much more precise
than any other observables, fitting data effectively gives $\,\widetilde{\delta \alpha}\simeq 0\,$.\,
In this sense, the parameter shift $\,\delta \alpha\,$ is always connected
to the dimension-6 operator coefficients. Nevertheless, we keep it free at the
moment, to give a general expression.

\vspace*{2mm}
\subsubsection{Fermi Constant}
\vspace*{1mm}

The Fermi constant is modified by the operators  $\mathcal O^{(3)}_L$  and
$\mathcal O^{(3)}_{LL}$ in \gtab{tab:O},
where the latter contributes a contact four-fermion vertex. Thus, we have
$\,\overline{\delta \GF}/\GF = 2 (v^2/\Lambda^2)(c^{(3)}_{LL} - c^{(3)}_L)$\,.\,
On the other hand, the effect of the $W$ field redefinition \geqn{eq:W-redef} is
cancelled by the correction \geqn{eq:W-del-mass} to its mass. Including the parameter shift,
we deduce the total effect,
\begin{equation}
  \frac {\widetilde{\delta \GF}}{\GF}
\,\simeq\,
  \frac {\delta \GF}{\GF}
+ 0.121
\left(
  \frac{c^{(3)}_{LL}}{\LambdaTeV^2}
- \frac{c^{(3)}_L}{\LambdaTeV^2}
\right) \!.
\label{eq:del-GF-tilde}
\end{equation}

\vspace*{2mm}
\subsubsection{Weak Gauge Boson Masses $\MW$ and $\MZ$}
\vspace*{1mm}

The contributions of dimension-6 operators to the $(W,Z)$ masses
have been summarized in Eqs.\geqn{eq:W-del-mass} and \geqn{eq:Z-del-mass}.
Including the parameter shifts, we derive the total contributions,
\begin{subequations}
\begin{eqnarray}
  \frac {\widetilde{\delta \MW}}{\MW}
& \simeq &
  0.184 \frac {\delta \GF}{\GF}
+ 1.37 \frac {\delta \MZ}{\MZ}
- 0.184 \frac {\delta \alpha} \alpha
+ 0.0262 \frac {c_{WW}^{}}{\LambdaTeV^2} \,,
\label{eq:del-MW-tilde}
\\[2mm]
  \frac {\widetilde{\delta \MZ}}{\MZ}
& \simeq &
  \frac {\delta M_Z}{M_Z}
- 0.0303 \frac {c_T^{}}{\LambdaTeV^2}
+ 0.0206 \frac {c_{WW}^{}}{\LambdaTeV^2}
+ 0.00149 \frac {c_{BB}^{}}{\LambdaTeV^2}
+ 0.00555 \frac {c_{WB}^{}}{\LambdaTeV^2} \,,~~~~~~
\label{eq:del-MZ-tilde}
\end{eqnarray}
\end{subequations}
for $W$ and $Z$ boson masses, respectively.

To make a consistent fit with the existing data, it is necessary to included the
SM radiative corrections. The coefficients of dimension-6 operators belong to the
next-to-leading (NLO) order. Up to the linear order of these NLO coefficients,
their contributions are independent of the SM loop corrections. Hence, the radiative
correction can be computed fully within the SM without involving new ultraviolet
divergence. Among the four observables $(\alpha, \GF, \MZ, \MW)$,
three of them can be used to fix the renormalization conditions, while
the remaining one receives a constant correction term. For convenience, we follow
the convention in \cite{Awramik:2003rn} by imposing renormalization conditions on
the SM predictions of $(\alpha,\,\GF,\,\MZ )$. Then, up to two-loop level, the $W$
mass becomes \cite{Awramik:2003rn},
\begin{equation}
  M^2_W \,=\,  M^2_Z
\left\{
  \frac 1 2 +
  \sqrt{\frac 1 4 - \frac {\pi \alpha}{\,\sqrt 2 \GF M^2_Z\,} \left[ 1 + \Delta r \right]\,}
\right\} .
\label{eq:mW2-a}
\end{equation}
The contribution of radiative corrections is included in $\,\Delta r\,$,\, which
is a function of electroweak parameters, $(\alpha,\,\GF,\,\MZ)$,\,
as well as Higgs mass $M_h^{}$ and the top quark mass $\,M_t$\,.\,
Since $\,\Delta r\,$ is already suppressed by loop factors,
the effect of varying its arguments is fairly small and
negligible up to the linear order. So $\,\Delta r\,$ can
be treated as a constant. For convenience, we define,
$\,\Delta r \equiv
  \Delta r_1^{} + \Delta r_2^{}$,\,
with $\,\Delta r_1^{}$ ($\Delta r_2^{}$)\,
denoting one-loop (two-loop) contributions.
For $\,M_h^{} = 125\,\mbox{GeV}$,\,
the values of $\Delta r_1^{}$ and $\Delta r_2^{}$
can be inferred from the Table\,1 of \cite{Awramik:2003rn},
$\,\Delta r_1^{} = 290.24 \!\times\! 10^{-4}$\, and
$\,\Delta r_2^{} = 72.99 \!\times\! 10^{-4}$.\,
The parameters $(\alpha,\,\GF,\,\MZ )$ have been precisely measured, with precision
much better than $10^{-4}$,\, while the radiative corrections
$\,\Delta r_1^{}\simeq 4\Delta r_2^{}=O(10^{-2})$.\,
So it is a reasonable approximation to
expand the corrected $W$ boson mass \geqn{eq:mW2-a} up to the linear order of
$\,\delta\alpha$, $\delta \GF$, $\delta \MZ$, $\Delta r_2^{}$, and the second order of
$\Delta r_1^{}$,
\begin{eqnarray}
  M_W^{} \!
= \!
  M^{(r)}_W \!
\left\{ \!
  1 \!
+ \!
  \frac 1 {\cos\! 2 \theta_w} \!
\left[
  c^2_w \frac {\delta M_Z}{M_Z}
+ \frac {s^2_w} 2
  \!\left(
    \frac {\delta G_F}{G_F} \!
  - \! \frac {\delta \alpha} \alpha
  \right) \!
- \!
  \frac {s^2_w} 2 \Delta r \!
- \!
  \frac {s^4_w (5 c^2_w \! - \! s^2_w)}{8 (c^2_w \! - \! s^2_w)^2} \Delta r^2_1
\right]
\right\} \!. \quad~~~
\end{eqnarray}
The dependence on the shifts of electroweak parameters remains the same as in
Eq.\,\geqn{eq:W-del-mass}. This is a general feature for any observables.
Loop corrections do not change the dependence on the shifts of electroweak
parameters up to the linear order, and only contribute as a constant term to
the observables.
By setting the reference values be the experimental central values \cite{PDG14},
$\,
  \alpha^{(r)} =
  7.297 352 5698 \!\times\! 10^{-3}$,\,
$  G^{(r)}_F =
  1.166 378 7 \!\times\! 10^{-5} \mbox{GeV}^{-2}$,\,
and
$\,M^{(r)}_Z = 91.1876\,\mbox{GeV}$,\,
the $W$ boson mass is predicted as
$\,M_W^{} = 80.385\,\mbox{GeV}$,\, which equals the
current experimental central value \cite{PDG14}.

Two remarks are in order.
First, in the above discussion we have imposed the renormalization conditions
on $(\alpha,\, \GF,\, \MZ)$.\, But one is free to choose any other renormalization
conditions. The difference caused by using different sets of renormalization conditions
only appears at higher order and thus can be ignored at the linear order analysis.
Second, since the dependence on parameter shifts
$(\delta\alpha,\, \delta G_F^{},\, \delta M_Z^{})$
remains the same as going from tree-level to one-loop level and the loop
corrections only contribute a constant term, all the expansions derived earlier
will continue to hold.

\vspace*{2mm}
\subsection{New Physics Contributions to Higgs Observables at $e^+ e^-$ Colliders}
\label{sec:collider}
\vspace*{1mm}

At future $e^+ e^-$ colliders
(such as the CEPC \cite{CEPC}, FCC-ee \cite{FCC-ee}, and ILC \cite{ILC}),
both productions and decays of the Higgs boson
can be systematically studied. The Higgs boson with mass
$\,M_h^{} = 125\,\mbox{GeV}$\,
is an ideal case for precision measurement of Higgs decay.
If $\,M_h^{}$\, would be either lighter or heavier than 125\,GeV,
the branching fractions would decrease very fast for some decay channels
($\,h \rightarrow WW, ZZ$\, when $\,h\,$ is too light, or
 $\,h \rightarrow \gamma\gamma,\,gg,\,f\bar{f}\,$ when $\,h\,$ is too heavy).
With $10^6$ Higgs bosons
to be collected at the CEPC, the Higgs decay into all gauge bosons and fermions
$\,(b,\, c,\, \tau,\, \mu)$\, can be measured. Both production and decay rates can help
to measure the Higgs coupling with other SM particles. The projected precision
of measuring the SM Higgs couplings can be extracted as we will elaborate in \gapp{app:cepc}.
In this subsection, we derive the corrections to these processes from new physics
as parametrized by the dimension-6 operators in \gtab{tab:O}.

\vspace*{2mm}
\subsubsection{Higgsstrahlung: $e^+e^-\!\to Zh$}
\label{sec:Zh}
\vspace*{1mm}

The Higgsstrahlung process  $\,e^+e^-\!\to Zh\,$
is the major production mode of the Higgs boson $h$\,(125GeV) at the Higgs factory
with center-of-mass energy $\,\sqrt s = 240-250\,\mbox{GeV}$.\, Its key advantage
is using the recoil mass distribution to make inclusive measurements, regardless of what
final-states the Higgs boson decays into.  The Higgs event rate can reach about $10^6$
at CEPC\,(250\,GeV) with an integrated luminosity of 5ab$^{-1}$ \cite{Ruan}.
From naive expectation, this cross section could be measured to
a precision level about
$\,\delta N/N \approx 1/\sqrt N = 0.1\%\,$.\,
The recent CEPC detector simulations \cite{CEPC} give the estimated sensitivity,
$\delta \sigma / \sigma \simeq 0.51\%$,\, at 68\%C.L.
\begin{figure}[h!]
\vspace*{3mm}
\centering
\includegraphics[height=4cm,width=0.7\textwidth]{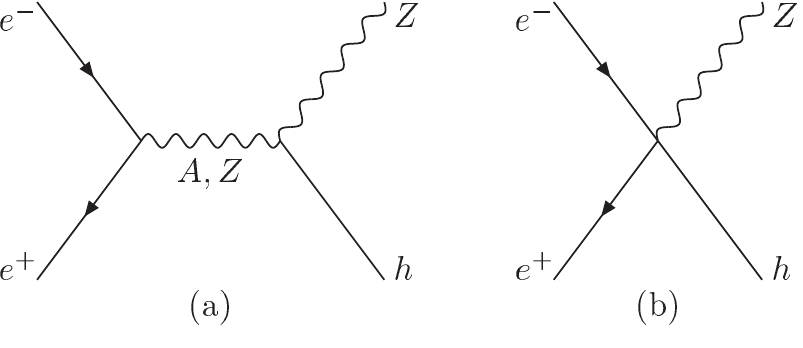}
\vspace*{-3mm}
\caption{Feynman diagrams for the Higgsstrahlung process
$\,e^+ e^-\!\rightarrow Z h$\,,\,
which include contributions of the relevant dimension-6 operators in Table\,\ref{tab:1}.
}
\vspace*{2mm}
\label{fig:Zh}
\label{fig:1}
\end{figure}

In \gfig{fig:Zh},
we summarize the relevant Feynman diagrams for $\,e^+ e^-\!\rightarrow Z h\,$ production,
which include possible contributions of the dimension-6 operators in \gtab{tab:O}.
Note that only the first diagram (a) has visible contributions, while other diagrams are
negligible due to the tiny Higgs-electron Yukawa coupling. This means that the Higgsstrahlung
is mainly mediated by $s$-channel gauge boson $Z^{}_\mu$ or $A^{}_\mu$.\,
The new physics contributions come
from corrections to vertices $\,Z \psi \bar \psi\,$ (cf.\ \gsec{sec:Zff}),
$ZZh$ and $AZh$ (cf.\ \gsec{sec:AZh}),
as well as $Zh \psi \bar \psi$ (cf.\ \gsec{sec:Zhff}). Among these,
the first does not introduce new topology since it contributes an
overall factor $\,\delta (g^2_L \!+\! g^2_R)$\, to the SM cross section.
This kind of contribution, including field redefinitions which can contribute to
the exiting vertices $Z \psi \bar \psi$ and $ZZh$, can be treated as a simple rescaling.
The others will either modify the tensor structure of the existing vertex
or introduce new vertex. We have systematically derived these contributions for the present study.
Since the final state consists of only the on-shell particles $Zh$,
we can present the results in analytical form.
We express the total cross section as a linear combination of the SM contribution
and the corrections of dimension-6 operators,
\begin{subequations}
\beqa
\sigma(Zh) \,=\, \(1 + 2 c^Z_0\)\sigma_{\text{sm}}^{} \,+\, \sum_{\,j,V~} c_j^{V} \sigma_j^{V} \,,
\label{eq:SigmaZh}
\eeqa
where $\,c_0^Z\,$ is defined in Eq.\eqref{eq:del-gZZh}, and
$\,c_j^{V}=c_j^{Z},c_j^{A}\,$ are given by
Eqs.\eqref{eq:fr-ZZh} and \eqref{eq:fr-AZh}.
For Eq.\eqref{eq:SigmaZh}, we derive $\,\sigma_{\text{sm}}^{}\,$ and $\,\sigma_j^{V}\,$
as follows,
	\begin{eqnarray}
	\sigma_{\text{sm}}^{}
	&=&
	 \frac{(g^2_R + g^2_L) g_{hZZ}^2}{48 \pi  M_Z^2\sqrt{s}} \frac {P_Z(P_Z^2+3M_Z^2)}{(s - M^2_Z)^2} ,
  \label{eq:Zh-sigma-SM}
	\\[1.5mm]
	\sigma^Z_1
	&=&
\frac{(g^2_R + g^2_L) g_{hZZ}^2 P_Z E_Z}{8\pi (s - M^2_Z)(s - M^2_V)} ,
\hspace{18mm}
  \sigma^A_1
\,=\,
 - \frac{\,g^2_{ZZh} (g_R^{} + g_L^{}) e\,}
				{8 \pi  (s - M^2_Z) s}
  E_Z^{} P_Z^{} \,,~~~~~
	\\[1.5mm]
	\sigma^Z_2
	&=&
	- 2 M_Z^2 \sigma_{\text{sm}}^{}\, ,
	\\[1.5mm]
	\sigma^Z_3
	&=&
	-2 \,s\, \sigma_{\text{sm}}^{}\, ,
\qquad
\hspace{33mm}
  \sigma^A_3 \,=\,
  \frac{\,2(g_L^{} \!+\! g_R^{}) e\,}{g^2_L \!+\! g^2_R}
  {(s \!-\! M^2_Z)}
  \sigma_{\text{sm}}^{} \,,~~~~~
\\[1mm]
  \sigma'_Z
    &=&
  2 \frac{\,g_L^{}\delta f_L^{} \!+\! g_R^{} \delta f_R^{}\,}
         {(g^2_L \!+\! g^2_R) g_{ZZh}}^{}
  (s \!-\! M^2_Z)
  \sigma_{\text{sm}}^{} \,,
  \label{eq:sigma-Zh'}
	\end{eqnarray}
\end{subequations}
where $(E_Z^{},P_Z^{})$ denote (energy, $|$momentum$|$) of the final-state $Z$ boson,
and the coefficients \,($c^Z_j$,\, $c^A_j$)\, are defined in \geqn{eq:fr-ZZh}-\geqn{eq:del-gZZh}
as well as \geqn{eq:fr-AZh}.
The corrections to fermionic coupling appear in
$\,\delta g_L^{}\,$ and $\,\delta g_R^{}\,$,\,
as summarized in \gsec{sec:Zff}.
In the last equation \geqn{eq:sigma-Zh'}, the corrections
$\,\delta f_L^{} = g_z^{} v (c^{(3)}_L \!\!+ c_L^{})/\Lambda^2$\,
and $\,\delta f_R^{} = g_z^{} v c_R^{} / \Lambda^2$\,
are coupling constants of the effective $\,\bar e e Z h\,$ vertex discussed in \geqn{eq:qqZh}.
Combining everything, we derive the relative
corrections to the cross section $\,\sigma(Zh)$\,,
\begin{eqnarray}
  \frac {\widetilde{\delta \sigma}} \sigma
& \,\simeq\, &
2.34 \frac {\delta G_F^{}}{G_F^{}}
+ 5.51 \frac {\delta M_Z^{}}{M_Z^{}}
- 0.344 \frac {\delta \alpha} \alpha
- 0.0605 \frac {c_H^{}}{\LambdaTeV^2}
- 0.206 \frac {c_T^{}}{\LambdaTeV^2}
\nonumber
\\
& &
+ 0.338 \frac {c_{WW}^{}}{\LambdaTeV^2}
+ 0.0122 \frac {c_{BB}^{}}{\LambdaTeV^2}
+ 0.0682 \frac {c_{WB}^{}}{\LambdaTeV^2}
+ 0.0429 \frac {c_{HW}^{}}{\LambdaTeV^2}
+ 0.00315 \frac {c_{HB}^{}}{\LambdaTeV^2}
\nonumber
\\
& &
+ 1.02 \frac {c^{(3)}_L}{\LambdaTeV^2}
+ 1.02 \frac {c_L^{}}{\LambdaTeV^2}
- 0.755 \frac {c_R^{}}{\LambdaTeV^2} .
\label{eq:del-sigma-Zh}
\end{eqnarray}

Comparing the above with the Eq.(3.10) of Ref.\,\cite{Craig:2014una},
we can see that our coefficient $c_T^{}$ is much larger,
due to the fact that we use scheme-independent approach
instead of the $Z$-scheme. The essential difference between these two approaches is due to the fact
that in the $Z$-scheme, $(\alpha, \GF, \MZ)$ are fixed to the measured values.
To reproduce the $Z$-scheme result from our scheme-independent approach, we can simply set
$\,\widetilde{\delta \alpha}\,$ in \geqn{eq:alpha},
$\,\widetilde{\delta\GF}\,$ in \geqn{eq:del-GF-tilde},
and $\,\widetilde{\delta \MZ}\,$ in \geqn{eq:del-MZ-tilde} to be zero.
In this way, the parameter shifts $(\delta \alpha, \delta G_F, \delta M_Z)$
can be expressed in terms of dimension-6 operator
coefficients. Then, implement these expressions of the parameter shifts into
\geqn{eq:del-sigma-Zh}. After these operations, the coefficient of $c_T^{}$ becomes
$-0.0397$, and agrees well with the value $-0.04$ in Ref.\,\cite{Craig:2014una}.\footnote{We
thank Matthew McCullough for detailed discussion and confirmation of this comparison with
Ref.\,\cite{Craig:2014una}.}

\vspace*{2mm}
\subsubsection{WW Fusion:
$\boldsymbol{e^+e^-\!\to \nu\bar{\nu}h}$ at 250\,GeV and 350\,GeV}
\vspace*{1.5mm}

\begin{figure}[t]
\centering
\includegraphics[width=0.95\textwidth]{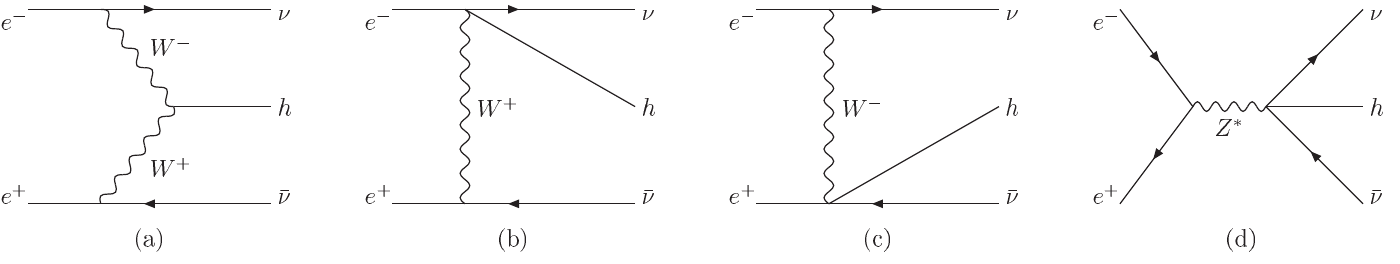}
\vspace*{-2mm}
\caption{Feynman diagrams for the $WW$ fusion process
$\,e^+ e^- \rightarrow h \nu \bar \nu$\,,\,
including contributions of the relevant dimension-6 operators in Table\,\ref{tab:1}.
}
\label{fig:nnh}
\label{fig:2}
\vspace*{3mm}
\end{figure}

The next production mode at the Higgs factory is
the $WW$ fusion process as depicted in \gfig{fig:nnh}.
Since the cross section $\sigma(\nu\bar{\nu}h)$ at $\sqrt{s}=250$\,GeV
is about $1/30$ of  $\,\sigma(Zh)$\, \cite{Ruan},
the $\sigma(\nu\bar{\nu}h)$
can be measured to a precision of $2.8\%$ at the CEPC \cite{CEPC}.
Although not as precise as the cross section $\sigma(Zh)$
of the Higgsstrahlung process, it can provide complementary
constraint on the Higgs coupling with $W$ gauge bosons.

The new physics contributions can be classified into two categories.
The first kind is the contribution to the
vertex $WWh$ with fusion topology, as studied in \gsec{sec:WWh}, which shares the same
Feynman diagram \gfig{fig:nnh}(a) as the SM contribution. Corresponding to the coefficients
$(c_0^W,c_1^{W},c_2^{W},c_3^{W})$ defined in \geqn{eq:cWs} and \geqn{eq:gWWh-del},
we derive the squared $S$-matrix elements,
\begin{subequations}
\begin{eqnarray}
  \overline{|\mathcal M|^2_0}
& = &
  \frac {4 g_{Wff}^4 g^2 M^2_W}
        {(M^2_W + 2 p_1^{}\! \cdot\! p_+^{})^2 (M^2_W + 2 p_2^{} \!\cdot\! p_-^{})^2}
  (p_1^{} \!\cdot\! p_-^{}) (p_2^{} \!\cdot\! p_+^{}) ,
\label{eq:WW-M0}
\\[2mm]
  \overline{|\mathcal M|^2_1}
& = &
\frac{-2 g_{Vff}^4 g^2 M^2_W (p_1^{}\!\cdot\! p_-^{}\!+p_2^{}\!\cdot\! p_+^{})}
  	 {(M^2_W\! + \! 2 p_1^{} \!\cdot\! p_+^{} \!)^2(M^2_W \! + \! 2 p_2^{} \!\cdot\! p_-^{} \!)^2}
  (2p_1^{}\!\cdot\! p_-^{} p_2^{}\!\cdot\! p_+^{}
   \!+2p_1^{}\!\cdot\! p_+^{} p_2^{}\!\cdot\! p_-^{} \!- s p_+^{}\!\cdot\! p_-^{}), ~~~~~~
\\[2mm]
  \overline{|\mathcal M|^2_2}
& \!\!=\!\!\! &
- 2 k^2_+ \overline{|\mathcal M|^2}_0
=
  4 (p_2^{} \!\cdot\! p_-^{}) \overline{|\mathcal M|^2_0} ~,
\\[2mm]
  \overline{|\mathcal M|^2_3}
& \!\!=\!\!\! &
- 2 k^2_- \overline{|\mathcal M|^2_0}
=
  4 (p_1^{} \!\cdot\! p_+^{}) \overline{|\mathcal M|^2_0} ~,
\end{eqnarray}
\end{subequations}
where $p_1^{}$ and $p_2^{}$ denote the momenta of $e^+$ and $e^-$,\, while
$p_+^{}$ and $p_-^{}$ are the momenta of $\nu$ and $\bar \nu$,\, respectively.
The zeroth-order term $\,\overline{|\mathcal M|^2_0}\,$ gives the SM contribution.
Only $\,\overline{|\mathcal M|^2_1}\,$ needs to be evaluated independently,
the rest are proportional to the zeroth-order result. For the first diagram,
its total effect is
$\,(1 \!+\! 2 c^W_0) \sigma_{\rm sm}^{} \!+ \sum_j c_j^{} \sigma_j^{}\,$,\,
with coefficients defined in Eqs.\,\geqn{eq:cWs} and \geqn{eq:gWWh-del}.
The second contribution comes from the new vertices in \gsec{sec:Whff}.
Their contribution to the $WW$ fusion is represented by the three new diagrams
\gfig{fig:nnh}(b)-(d),
\begin{subequations}
\begin{eqnarray}
  \delta|\mathcal M_{W h \ell \bar \nu}^{}|^2
& =\, &
  2|\mathcal M_0^{}|^2 \frac{g_{W h \ell \bar \nu}^{}}
  {\,g_{W e \nu}^{}g_{hWW}^{}\,}
  \left[ (k^2_+\! - M^2_W) + (k^2_-\! - M^2_W) \right] ,
\\[2mm]
  \delta |\mathcal M_{Zh \nu \bar \nu}^{}|^2
& =\, &
  2|\mathcal{M}_{0}^{}|^2
  \frac{\,(k^2_+\! - M^2_W)(k^2_-\! - M^2_W)\,}{s-M_z^2}
  \!\left(\!\!
  -\frac{\,g_{ZeeL}^{} g_{Z h \nu \bar \nu}^{}\,}{g_{We\nu}^2 g_{hWW}^{}}
  \!\right) ,~~~~~
\end{eqnarray}
\end{subequations}
where $\,k_\pm^{}\,$ denotes the momenta of $\,W^\pm$. Both contributions come from
the interference with the SM contribution $\mathcal M_0$. For convenience we have denoted
the couplings as $\,g_{W h \ell \bar \nu}^{} \equiv \sqrt 2 g v c^{(3)}_L / \Lambda^2$,\,
$g_{Zh \nu \bar \nu}^{} \equiv g_z^{} v(c^{(3)}_L \!- c_L^{}) / \Lambda^2$,\,
$g_{ZeeL}^{} \equiv g_z^{} ( \frac 1 2 \!- s^2_w)$,\,
$g_{W e \nu}^{} \equiv g/\!\sqrt 2$,\, and $\,g_{hWW}^{} \equiv g^2 v / 2$\,,\,
where $\,g_z^{} \equiv g/\cos \theta_w^{}\,$.\,
Note that only the left-handed part of the neutral current
$\,g_{Zee,L}^{} Z_\mu^{} \bar e_L^{} \gamma^\mu e_L^{}\,$
in \gfig{fig:nnh}(d) can interfere with the SM
contribution $\,\mathcal M_0^{}$\,.\, After combining the two contributions, we derive
\begin{subequations}
	\begin{eqnarray}
	250\,\mbox{GeV}\!: \dis\frac{\widetilde{\delta \sigma}} \sigma
	& \,\simeq\, &
	3.44 \frac {\delta G_F^{}}{G_F^{}}
	+ 3.28 \frac {\delta M_Z^{}}{M_Z^{}}
	- 0.442 \frac {\delta \alpha} \alpha
	- 0.0605 \frac {c_H^{}}{\LambdaTeV^2}
\nonumber
\\
& + &
	  0.0515 \frac {c_{WW}^{}}{\LambdaTeV^2}
	+ 0.0126 \frac {c_{HW}^{}}{\LambdaTeV^2}
	- 0.159 \frac {c^{(3)}_L}{\LambdaTeV^2}
	+ 0.0136 \frac {c_L}{\LambdaTeV^2} ,
\qquad
	\\[2mm]
	 350\,\mbox{GeV}\!: \dis\frac {\widetilde{\delta \sigma}} \sigma
	& \simeq\, &
	3.52 \frac {\delta \GF}{\GF}
	+ 3.89 \frac {\delta \MZ}{\MZ}
	- 0.523 \frac {\delta \alpha} \alpha
	- 0.0605 \frac {c_H^{}}{\LambdaTeV^2}
\nonumber
\\
& + &
	  0.0575 \frac {c_{WW}^{}}{\LambdaTeV^2}
	+ 0.0188 \frac {c_{HW}^{}}{\LambdaTeV^2}
	- 0.226 \frac {c^{(3)}_L}{\LambdaTeV^2}
	+ 0.00918 \frac {c_L}{\LambdaTeV^2} ,
	\end{eqnarray}
\label{eq:sigma-nnh}
\end{subequations}
\hspace*{-3mm}
for $\,\sqrt s = 250\,\mbox{GeV}$ and 350\,GeV, respectively.
At $\,\sqrt s = 350$\,GeV, we see that the Higgs production cross section through
$WW$ fusion has sizable increase, leading to a better
measurement of $\,\sigma(\nu \bar \nu h)$\,.

\vspace*{2mm}
\subsubsection{Higgs Decay into $Z$ Boson Pair}
\vspace*{1.5mm}

For Higgs decay into the $Z$ boson pair, at least one of them must be off-shell.
The decay width can be computed via the corresponding three-body decay process,
$\,h \rightarrow ZZ^* \rightarrow Z f \bar f$\,.\,
In addition, the double off-shell process $\,h\to Z^*Z^*$\, still contributes
$25\%$ of the partial width and thus should be included
via the four-body decay process,
$\,h \rightarrow Z^* Z^* \rightarrow f_1^{} \bar f_1^{} f_2^{} \bar f_2^{}$\,.\,
We compute the new physics contributions to the Higgs partial width
by using FeynRules \cite{FeynRules} and MadGraph5 \cite{Mad5}.
With these, we derive the following expression,
\begin{eqnarray}
\frac{\,\widetilde{\delta\Gamma}\,}{\Gamma}
& \simeq &
  3.42 \frac{\delta G_F^{}}{G_F^{}}
- 5.44 \frac{\delta M_Z^{}}{M_Z^{}}
- 0.420 \frac{\delta\alpha}{\alpha}
- 0.0605\frac{c_H^{}}{\LambdaTeV^2}
+ 0.190 \frac{c_T^{}}{\LambdaTeV^2}
\nonumber
\\
&&
- 0.0968 \frac {c_{WW}^{}}{\LambdaTeV^2}
- 0.0255 \frac {c_{BB}^{}}{\LambdaTeV^2}
- 0.0579 \frac {c_{WB}^{}}{\LambdaTeV^2}
+ 0.0131 \frac {c_{HW}^{}}{\LambdaTeV^2}
+ 0.0144 \frac {c_{HB}^{}}{\LambdaTeV^2}
\nonumber
\\
&&
+ 0.0410  \frac{c^{(3)}_L}{\LambdaTeV^2}
- 0.0112  \frac{c_L^{}}{\LambdaTeV^2}
- 0.00957 \frac{c_R^{}}{\LambdaTeV^2}
\nonumber
\\
&&
+ 0.101   \frac{c^{(3)}_{L,q}}{\LambdaTeV^2}
+ 0.0269  \frac{c_{L,q}^{}}{\LambdaTeV^2}
+ 0.0128  \frac{c_{R,u}^{}}{\LambdaTeV^2}
- 0.00957 \frac{c_{R,d}^{}}{\LambdaTeV^2} .
\label{eq:Gamma-hZZ}
\end{eqnarray}
The CEPC detector simulations \cite{CEPC} show that this decay branching fraction
can be measured to the precision of $4.3\%$\,.

\vspace*{2mm}
\subsubsection{Higgs Decay into $W$ Boson Pair}
\vspace*{1.5mm}

The analysis of this process is similar to that of $\,h\to ZZ$\,.\,
We use FeynRules \cite{FeynRules} and MadGraph5 \cite{Mad5} to numerically compute
the new physics contributions to $h\to WW^*,W^*W^*$
with 3-body and 4-body final states. Altogether, we derive the
contributions of the relevant dimension-6 operators to the following,
\begin{eqnarray}
  \frac {\widetilde{\delta \Gamma}} \Gamma
& \simeq &
  1.64 \frac {\delta G_F^{}}{G_F^{}}
- 10.1 \frac {\delta M_Z^{}}{M_Z^{}}
+ 1.36 \frac {\delta \alpha} \alpha
- 0.0605 \frac {c_H^{}}{\LambdaTeV^2}
\nonumber
\\
&&
- 0.233 \frac {c_{WW}^{}}{\LambdaTeV^2}
+ 0.0225 \frac {c_{HW}^{}}{\LambdaTeV^2}
+ 0.0479 \frac {c^{(3)}_L}{\LambdaTeV^2}
+ 0.0968 \frac {c^{(3)}_{L,q}}{\LambdaTeV^2} .
\label{eq:Gamma-hWW}
\end{eqnarray}
The branching fraction of $\,h \rightarrow WW\,$ can be measured with to
$\,1.5\%\,$ accuracy at the CEPC \cite{CEPC}. Note that this channel is
measured with better precision than $\,h \rightarrow ZZ$\,.\, This is because
$W$ is lighter than $Z$, and hence the $\,WW\,$ channel has much larger
branching fraction than the $ZZ$ channel.
This difference in decay rates leads to different precisions which are mainly
dominated by statistical fluctuations.

\vspace*{2mm}
\subsubsection{Other Decay Channels}
\vspace*{1.5mm}

The remaining Higgs decay channels can be divided into two major classes: one
with fermionic decay products and the other with massless gauge bosons (photons or gluons).
The first class occurs at tree-level, while the second class arises from one-loop level.
Both receive contributions from the Higgs field redefinition \geqn{eq:h-redef},
\begin{equation}
  \frac{\,\delta\Gamma\,}{\Gamma} \,=\,
- 0.0605 \frac {c_H^{}}{\,\LambdaTeV^2\,}\, ,
\end{equation}
which is the only contribution to fermionic decays.
The vertex $\,f\bar{f}h\,$ comes from Yukawa interaction
which flips chirality and is not affected by either the dimension-6
operators in \gtab{tab:O} or the EW parameters mentioned earlier.
On the other hand, the decay into photons has extra contributions.
For fermion loop, it is affected by the photon
field redefinition \geqn{eq:A-redef} only.
For bosonic $W$-loop, the new physics effects come from
$W$-mass correction \geqn{eq:W-del-mass} and the photon field redefinition \geqn{eq:A-redef}.
Note that the corrections of $W$-field redefinition
to the vertex and mass should cancel with each other. Since the EW
parameters are involved in bosonic decay,
their shifts $(\delta \alpha,\, \delta \GF,\, \delta\MZ)$ should also appear.
Note that $\,h \rightarrow gg\,$ only has fermionic contributions.

Furthermore, dimension-6 operators induce direct coupling of the Higgs field $h$
with photons or gluons, as shown in Eqs.\,\geqn{eq:FR-hAA} and \geqn{eq:FR-hgg},
respectively. Thus, we derive the following $hAA$ and $hgg$ couplings,
\begin{subequations}
\begin{eqnarray}
\label{eq:hgagaM}
	\mathcal M_{hAA}^{}
& = &
  \frac{4}{\,v\,} \delta Z_A^{}
  f^{\mu\nu} \llbracket p_1^{}, p_2^{} \rrbracket
  \epsilon_{1\mu}^{} \epsilon_{2\nu}^{} \,,
  ~~~~~
\\[1mm]
  \mathcal M_{hgg}^{}
& = &
  \frac{4v}{\,\Lambda^2\,} g_s^2 c_g^{}
  f^{\mu \nu} \llbracket p_1^{}, p_2^{} \rrbracket
  \,\delta_{ab}^{}\, \epsilon_{1\mu}^{} \epsilon_{2\nu}^{} \,.
\end{eqnarray}
\label{eq:hAG}
\end{subequations}
We may compare them with the corresponding SM-loop results,
\begin{subequations}
\begin{eqnarray}
\label{eq:hgagaMsm}
  \mathcal M_{hAA}^{\text{sm}}
& \,=\, &
  \frac{e^2}{\,8\pi^2 v\,}\(F_W+\sum N_c Q_f^2 F_f\)\!
  f^{\mu \nu} \llbracket p_1, p_2 \rrbracket \epsilon_{1\mu} \epsilon_{2\nu}  ,
\\[1mm]
\label{eq:hggMsm}
  \mathcal M_{hgg}^{\text{sm}}
& \,=\, &
  \frac {\alpha_s} {\,4\pi v\,}  F_f^{} f^{\mu \nu}
  \llbracket p_1^{}, p_2^{} \rrbracket
  \,\delta_{ab}^{}\, \epsilon_{1\mu}^{} \epsilon_{2\nu}^{} ,
\\[2mm]
&&
F_W^{} \equiv~ 2+3 \tau_W^{-1} \left[1 + (2- \tau_W^{-1}) f(\tau_W^{})\right]\!,
\\[1mm]
&&
F_f^{} ~\equiv\, - 2 \tau_f^{-1} \left[1 + (1- \tau_f^{-1}) f(\tau_f^{})\right]\!,
\eeqa
\label{eq:hAG-SM}
\end{subequations}
where
$\,\tau_j^{} \equiv {M_h^2}/{(4 M_j^2)}$\, and
$\,f(\tau_j^{}) \equiv (\arcsin\!\sqrt{\tau_j^{}})^2\,$,\,
with $\,j = W, t$.\,
Combining everything together, we derive the total corrections,
\begin{subequations}
\begin{eqnarray}
\frac{\,\delta \Gamma_{AA}^{}\,}{\Gamma_{AA}^{}}
& = &
  0.997 \frac {\delta \GF}{\GF}
+ 2 \frac {\delta \alpha}{\alpha}
- 0.0218 \frac{\delta\MZ}{\MZ}
- 0.0605 \frac{c_H^{}}{\,\LambdaTeV^2}
+ 5.91 \frac{\,c_{WW}^{} \! - \! c_{WB}^{} \! + \! c_{BB}^{}\,}{\LambdaTeV^2} ,
~~~~~~\qquad
\label{eq:dGammaAA}
\\[2mm]
  \frac{\,\delta\Gamma_{gg}^{}\,}{\Gamma_{gg}^{}}
& = &
  \frac{\delta\GF}{\GF}
- 0.0605 \frac {c_H^{}}{\,\LambdaTeV^2}
- 55.2 \frac {c_g^{}}{\,\LambdaTeV^2} \,,
\label{eq:dGammaGG}
\end{eqnarray}
\end{subequations}
for $\,\Gamma_{AA}^{}\,$ and $\,\Gamma_{gg}^{}$\,,\, respectively.
We note that the coefficients of the last terms in both \geqn{eq:dGammaAA}
and \geqn{eq:dGammaGG} come from the interference between the SM
prediction \geqn{eq:hAG-SM} and the contribution \geqn{eq:hAG}
by dimension-6 operators. Although the SM predictions of
$\,h\to gg\,$ and $\,h\to\gamma\gamma\,$
arise from loop-level and are expected to be of the same order
as that of dimension-6 operators, it is well justified to make expansion up to
the linear terms of $\,c_g^{}$\, and $\,(c_{WW}^{},c_{WB}^{},c_{BB}^{})$.\,  This
is because the current LHC data constrain the deviations from the SM predictions
within about 20\% at $2\sigma$ level \cite{LHC-fit},
and the future Higgs factory sensitivities
to such deviations are even much smaller
(Table\,\ref{tab:2} in Sec.\,\ref{sec:4.3} and Fig.\,\ref{fig:5} in Appendix\,\ref{app:C}).
Hence, the dimension-6 contributions
can be well treated as small perturbations up to the linear order.

\vspace*{2mm}
\subsection{Probing New Physics Scales at Higgs Factory}
\label{sec:scale}
\label{sec:4.3}
\vspace*{1.5mm}

As discussed in \gsec{sec:precision} and \gsec{sec:collider}, the dimension-6
effective operators can modify both EW precision observables (EWPO)
and Higgs observables (HO).
The EWPO have been precisely measured at the LEP and Tevatron with high precision,
while the HO can be measured at the future Higgs factory under planning.
Currently, there are three major candidates of Higgs factory,
CEPC \cite{CEPC}, FCC-ee \cite{FCC-ee}, and ILC \cite{ILC}, which can run at
the collision energies around $240-250$\,GeV.
They can measure the Higgs production
cross sections and decay branching fractions with precisions at percentage level.
This provides important means to indirectly probe the scales of new physics.
In the following, we study how the EWPO and HO
can probe the new physics scales via effective dimension-6 operators
and the interplay with each other.

For convenience, we first summarize the inputs for our analysis in \gtab{tab:inputs}.
Since the EWPO have already been measured, we list both their central values
and relative errors. These four observables are the most precisely
measured ones. Especially, the fine-structure constant $\alpha$ is measured with
unprecedented precision of
$\,\delta\alpha/\alpha = 3.29 \!\times\! 10^{-10}$,\,
much better than all the others.
According to its expression \geqn{eq:alpha}, one degrees of freedom can be
effectively eliminated. This is also true for the Fermi constant $\,\GF$\,,\, whose
precision  $\,\delta\GF/\GF = 5.14 \!\times\! 10^{-7}$\, is just next to
that of $\,\alpha$\,.

\begin{table}[t]
\centering
\caption{Inputs used to constrain the new physics scales of
         the dimension-6 operators. The electroweak precision observables
         in the first four rows are taken from PDG \cite{PDG14},
         and the estimated precisions of Higgs measurements (68\%\,C.L.) are given by
         the CEPC detector simulations \cite{CEPC}.
         For the $WW$ fusion cross section $\sigma[\nu \bar \nu h]_{350\text{GeV}}^{}$
         at $\sqrt{s}=350$\,GeV, we adopt the
         FCC-ee (TLEP) estimation \cite{FCC-ee} for illustration.
         For the ``Measurements'' entry,
         the number inside the parentheses stands for experimental uncertainty.}
\vspace*{3mm}
\begin{tabular}{c||cc|c}
\hline \hline
Observables & Measurements & Relative Error & SM Prediction
\\
\hline \hline
$\MZ$ & 91.1876(21)\,GeV & $2.3 \times 10^{-5}$ & --
\\
$\MW$ & 80.385(15)\,GeV  & $1.87 \times 10^{-4}$ & --
\\
$\GF$ & $1.166  378  7(6) \!\times\! 10^{-5} \mbox{GeV}^{-2}$ & $5.14 \!\times\! 10^{-7}$ & --
\\
$\alpha$ & $7.297  352  569  8(24) \!\times\! 10^{-3}$ & $3.29 \!\times\! 10^{-10}$ & --
\\
\hline
$\sigma[Zh]$ & -- & 0.50\% & -- \\
$\sigma[\nu \bar \nu h]$ & -- & 2.86\% & -- \\
{~~~~~~~~~$\sigma[\nu \bar \nu h]_{350\text{GeV}}^{}$} & -- & {0.75\%} & -- \\
\hline
$\text{Br}[WW]$ & -- & 1.2\% & 22.5\% \\
$\text{Br}[ZZ]$ & -- & 4.3\% & 2.77\% \\
$\text{Br}[bb]$ & -- & 0.54\% & 58.1\% \\
$\text{Br}[cc]$ & -- & 2.5\% & 2.10\% \\
$\text{Br}[gg]$ & -- & 1.4\% & 7.40\% \\
$\text{Br}[\tau \tau]$ & -- & 1.1\% & 6.64\% \\
$\text{Br}[\gamma \gamma]$ & -- & 9.0\% & 0.243\% \\
$\text{Br}[\mu \mu]$ & -- & 17\% & 0.023\% \\
\hline \hline
\end{tabular}
\label{tab:inputs}
\label{tab:2}
\vspace*{2mm}
\end{table}

For the Higgs observables, Table\,\ref{tab:2} summarizes the
estimated precisions at the CEPC \cite{CEPC}.
The production cross sections and branching fractions are independent of each other.
Nevertheless, the decay widths \geqn{eq:Gamma-hZZ}-\geqn{eq:Gamma-hWW} for
Higgs decays into $ZZ$ and $WW$ bosons cannot be directly used to
compare with the branching fraction precisions in \gtab{tab:inputs}. The decay width
for a specific channel competes with all other channels, so its corresponding
branching fraction is given by
$\,\text{Br}_j^{} \equiv \Gamma_j^{} / \Gamma$,\, where
$\,\Gamma \equiv \sum_k \Gamma_k^{}$\,
is the total decay width. Each partial width can be expressed as
$\,\Gamma_j^{} \equiv \Gamma^{(r)}_j (1 + \delta \Gamma_j^{} / \Gamma_j^{})$\,
with $\,\delta \Gamma_j^{}\,$ denoting the deviation from the reference point.
When expanded to linear order, the decay branching fraction becomes,
\begin{equation}
  \text{Br}_j^{}
~\simeq~
  \text{Br}^{(r)}_j\!
\left[
  1
+ \left( 1 - \text{Br}^{(r)}_j \right)\! \frac{\,\delta\Gamma_j^{}\,}{\Gamma_j^{}}
- \sum_{k \neq j} \text{Br}^{(r)}_k \frac{\,\delta \Gamma_k^{}\,}{\Gamma_k^{}}
\right] \!.
\label{eq:Br-i}
\end{equation}
The corrections to Higgs partial width affect not only its own branching fraction,
but also all the others.
Eq.\geqn{eq:Br-i} shows that
for the branching fraction $\,\text{Br}_j^{}$,\, the contribution due to its own channel
is modulated by $\,1\! - \text{Br}^{(r)}_j$,\, while other channels by the corresponding
$\,\text{Br}^{(r)}_k$.\,  Since the reference value is around
the SM prediction, $\,\text{Br}^{(r)}_j \!\approx \text{Br}^\text{sm}_j$,\,
the modulation is essentially controlled by the SM predictions.
In this way, the precision measurements of branching fractions at CEPC
will constrain the new physics scales via
$\,\delta \Gamma_j^{}\,$ term and
$\,\delta \Gamma_k^{}\,$ term.

The observables in \gtab{tab:inputs} can be used to constrain the electroweak parameters
$(\delta \alpha$, $\delta \GF$, $\delta \MZ)$ and the coefficients of dimension-6 operators
simultaneously. This can be achieved by the so-called $\chi^2$ fit technique. As described in
Appendix\,\ref{sec:chi2}, the $\chi^2$ function sums over all experimental observables
$\,\mathcal O_j^{}$\,,
\begin{equation}
  \chi^2 \!\left( \delta\alpha, \delta\GF, \delta\MZ, \frac {c_i^{}}{\Lambda^2} \right)
~=~
\sum_j \left[
\frac{\,\mathcal O^{\th}_j\! \left( \delta \alpha, \delta \GF,
\delta \MZ, \frac {c_i^{}}{\Lambda^2} \right) - \mathcal O^{\exp}_j\,}
{\Delta \mathcal O_j} \right]^{\!2} ,~~~~~
\end{equation}
where the theoretical predictions are functions of the fitting parameters. The $\chi^2$ function
reaches its minimal value at the best fit values
of $(\delta\alpha, \delta\GF, \delta\MZ)$ and $c_i^{}/\Lambda^2$. Using the
linear $\chi^2$ fitting method shown in Appendix\,\,\ref{sec:chi2},
we can perform this fit analytically with the package BSMfitter\,\cite{BSMfitter}.
As usual, for simplicity, we will consider only one dimension-6 effective operator
to be nonzero during each fit, and turn off the others.
Thus, each fit will deal with only four fitting parameters,
$(\delta \alpha, \delta\GF, \delta\MZ)$ and one dimension-6 coefficient
$\,c_i^{}/\Lambda^2\,$.

\begin{figure}[t]
\centering
\includegraphics[height=9cm]{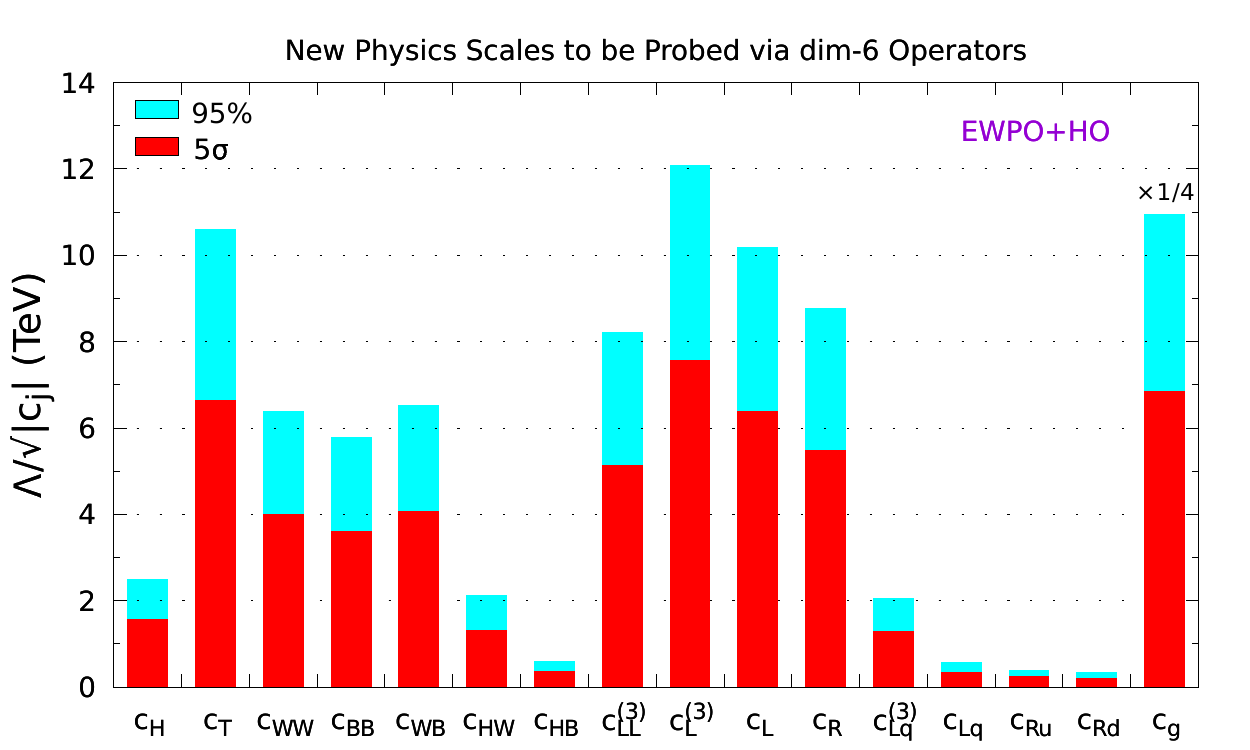}
\caption{The 95\% exclusion limits (blue) and $5\sigma$ discovery sensitivities (red) to
the new physics scales $\,\Lambda/\!\sqrt{|c_j|}\,$
by combining the current electroweak precision observables
($\alpha, \,\GF, \,\MZ,\, \MW$) \cite{PDG14} and the future Higgs observables
(Table\,\ref{tab:2}) at the Higgs factory CEPC\,(250\,GeV) \cite{CEPC}
with a projected luminosity of \,5\,ab$^{-1}$.
In the last column for $\mathcal O_g^{}$, we have rescaled its height
by a factor $1/4$ to fit the plot, so its actual reach is
$\,\Lambda /\!\sqrt{|c_g|}= 43.8\,$TeV.
}
\label{fig:NP-scale}
\label{fig:3}
\end{figure}

In \gfig{fig:NP-scale}, we present the lower limit
on the new physics scale
of each dimension-6 operator by combining the existing electroweak precision
measurements and future Higgs measurements at the CEPC with
$\sqrt s = 250\,\mbox{GeV}$. We see that it can
probe the new physics scales up to about $12\,\mbox{TeV}$ for $\mathcal O^{(3)}_L$
at 95\%\,C.L.

For the operators listed in \gtab{tab:O},  ($\mathcal O_T$,
$\mathcal O^{(3)}_{LL}$, $\mathcal O^{(3)}_L$, $\mathcal O_{L,R}^{}$,
$\mathcal O_g^{}$) are among the first group to be sensitively probed.
Roughly speaking, they can be probed up to the new physics scales
$\,(8-10)\,\mbox{TeV}$.\,  The second group consists
($\mathcal O_H^{}$, $\mathcal O_{WW}^{}$, $\mathcal O_{BB}^{}$,
$\mathcal O_{WB}^{}$, $\mathcal O_{HW}^{}$, $\mathcal O^{(3)}_{Lq}$), which
can be probed up to the scales $\,(2-5)\,\mbox{TeV}$.\, The others operators,
($\mathcal O_{HB}^{}$, $\mathcal O_{Lq}^{}$, $\mathcal O_{Ru}^{}$, $\mathcal O_{Rd}^{}$),
cannot be probed above the \,1\,TeV scale.  We note that the strong
constraint on $\mathcal O_T^{}$ mainly comes from the $W$ boson mass $\MW$.\,
Including electroweak precision observables can significantly improve the probe of new physics
scales, as we will fully elaborate in following
\gsec{sec:EWPO} and \gsec{sec:ZWmass}. The remaining constraints come
from measuring the Higgs production and decay rates, most of which is provided by
the Higgsstrahlung process. For the gluonic operator $\,\mathcal O_g^{}$,\,
its constraint is mainly given by the branching fraction of $\,h \rightarrow gg\,$
which is the only relevant channel here.
Although this is not the major Higgs decay channel,
with the SM prediction $\,\text{Br}[gg] = 7.4\%$,\,
it can put severe constraint on the scale of $\,\mathcal O_g^{}$,\,
as high as about 43.8\,TeV (cf.\ \gfig{fig:3}).
Since in the SM the Higgs coupling with gluons arises at one-loop level
and the dimension-6 operator $\,\mathcal O_g^{}$\, contributes to this coupling
at tree-level, so the scale of $\,\mathcal O_g^{}$\, has to be high enough
to suppress the deviation from the SM loop prediction.
This is expected since the operator $\,\mathcal O_g^{}$\, may well be induced
from loop-level in a given underlying theory and thus its coefficient $\,c_g^{}\,$
will be suppressed by the corresponding loop-factor.

Note that \gfig{fig:3} contains more fermionic operators than listed in
\gtab{tab:O} since quark and lepton can provide different contributions.
For a specific operator, we assume the same operator coefficient for the
three generations of fermions. Consequently, each of the operators involving left-handed
fermions, ($\mathcal O^{(3)}_{LL},\, \mathcal O^{(3)}_L,\, \mathcal O_L^{}$),
has two copies, one for leptons and the other for quarks (with
extra subscript ``$q$''). On the other hand, the operator $\,\mathcal O_R^{}\,$ that
contains the right-handed fermions has three copies, one for charged leptons
and the other two for quarks (with subscripts ``$u$'' for up- and ``$d$\,'' for down-type
quarks). We can see that leptonic operators are generally better
constrained than those of quarks, since the former can enter the most
precisely measured Higgsstrahlung process, and the latter can only be constrained
by Higgs decays into $WW$ and $ZZ$ with limited branching fractions and statistics.
Although the Higgs decay mode $\,h \rightarrow b \bar b\,$
has the largest branching fraction, it is not connected to the fermionic operators
shown in \gtab{tab:O}.

\begin{table}[t]
\vspace*{3mm}
\setlength{\tabcolsep}{0.7mm}
\centering
\caption{New physics scales $\Lambda/\!\sqrt{|c_j|}$ (in TeV) which can be probed by combining
         the current electroweak precision tests on ($\alpha, \GF, \MZ, \MW$) \cite{PDG14}
         and the future Higgs measurements on ($\sigma(Zh)$, $\sigma(\nu \bar \nu h)$, and
         branching fractions) at the Higgs factory CEPC\,(250\,GeV) \cite{CEPC} with a projected
         luminosity of 5\,ab$^{-1}$. The sensitivities are presented as the $95\%$
         exclusions (first row) and the $5\,\sigma$ discoveries (second row), respectively.}
\vspace*{3mm}
\begin{tabular}{cccccccccccccccc}
\hline \hline
$\mathcal O_H^{}$ & $\mathcal O_T^{}$ & $\mathcal O_{WW}^{}$
& $\mathcal O_{BB}^{}$ & $\mathcal O_{WB}^{}$
& $\mathcal O_{HW}^{}$ & $\mathcal O_{HB}^{}$
& $\mathcal O^{(3)}_{LL}$ & $\mathcal O^{(3)}_L$
& $\mathcal O_L^{}$ & $\mathcal O_R^{}$
& $\mathcal O^{(3)}_{L,q}$ & $\mathcal O_{L,q}^{}$
& $\mathcal O_{R,u}^{}$ & $\mathcal O_{R,d}^{}$ & $\mathcal O_g^{}$
\\
\hline
  2.5 & 10.6 & 6.38 & 5.78 & 6.53 & 2.12 & 0.604 & 8.23 & 12.1 & 10.2 & 8.78 & 2.06 & 0.568 & 0.393 & 0.339 & 43.8 \\
  1.57 & 6.65 & 4.00 & 3.62 & 4.09 & 1.33 & 0.378 & 5.15 & 7.57 & 6.39 & 5.49 & 1.29 & 0.356 & 0.246 & 0.212 & 27.4
\\
\hline \hline
\end{tabular}
\label{tab:scale}
\label{tab:3}
\end{table}

For clarity, in \gtab{tab:scale}, we further present the numerical limits
of \gfig{fig:NP-scale} at both 95\% and $5\sigma$ confidence
levels. The 95\% limit corresponds to the exclusion reach, while $5\sigma$ limits
gives the discovery reach. Since the results are obtained after reducing to one-dimensional
Gaussian distribution by marginalization (see \gapp{sec:chi2} for detail),
the value of the $5\sigma$ reach on the new physics scale equals 39\%
of the corresponding 95\% confidence limit.

Note that these results are obtained with all Higgs observables to be measured
at $\sqrt s = 250\,\mbox{GeV}$. If the collision energy is upgraded to $350\,\mbox{GeV}$,
the cross section of the $WW$ fusion process for Higgs production will increase
significantly. This can help to enhance the sensitivity to the scale of $\,\mathcal O_H^{}\,$ by
about \,10\%,\, as will be shown in the first column of \gtab{tab:ZWmass-effects},
while the others remain the same.

\vspace*{2mm}
\subsection{Combining with Electroweak Precision Observables}
\label{sec:EWPO}
\vspace*{1.5mm}

For comparison, we note that the $Z$-scheme is adopted
in the recent studies \cite{Craig:2014una} and \cite{Ellis:2015sca},
where the latter also invokes the $W$ mass measurement at a Higgs factory.
In this scheme, not all the electroweak parameters, especially the most precisely
measured ones ($\alpha, \GF, \MZ$), were included in their analysis.
After incorporating the electroweak precision measurements, including also
$\MW$,\, the reach of new physics scales\,\cite{Ellis:2015sca} becomes higher
than the one with the $\,\sigma(Zh)\,$ constraints alone \cite{Craig:2014una}.
Although the $\,\MW\,$ measurement is also used in \cite{Ellis:2015sca}, its interplay
with $\MZ$ could not be studied within the $Z$-scheme.
In this subsection, we first study the role of electroweak precision observables
(EWPO) with the current data\,\cite{PDG14}. We will further analyze the interplay of
including a significantly improved $\MW$  measurement in \gsec{sec:ZWmass}.

Among the existing EWPO, the most precisely measured observables
are $\alpha$, $\GF$, and $\MZ$, in the order of their relative uncertainties,
as shown in \gtab{tab:inputs}. Even the least precise one, $\MZ$,
is much better measured than the other mass $\MW$ by about one order of magnitude.
This hierarchical structure in the relative uncertainties makes it appropriate
to treat $(\alpha, \GF, \MZ)$ as inputs to fix the electroweak parameters ($g, g', v$),
and implement the $\MW$ measurement into the fit. As we discussed
in \gsec{sec:Zh}, this is equivalent to setting
$\,(\widetilde{\delta\alpha},\, \widetilde{\delta\GF},\, \widetilde{\delta \MZ})=0$\,,\,
from which $(\delta\alpha, \delta\GF, \delta\MZ)$ can
be solved in terms of dimension-6 operator contributions.
With these extra constraints (which is exactly the definition of $Z$-scheme)
implemented into \geqn{eq:del-MW-tilde}, we derive the $W$ mass correction as
\begin{equation}
  \frac {\widetilde{\delta M_W}}{M_W}
\,=~
  0.0414 \frac {c_T^{}}{\LambdaTeV^2}
- 0.00964 \frac {c_{WB}^{}}{\LambdaTeV^2}
- 0.0223 \frac {c^{(3)}_{LL}}{\LambdaTeV^2}
+ 0.0223 \frac {c^{(3)}_L}{\LambdaTeV^2} \,,~~~~~
\label{eq:del-MW-Zscheme}
\end{equation}
which is a function of the coefficients of dimension-6 operators alone.
In Eq.\,\eqref{eq:del-MW-Zscheme},
even though the coefficients of the $\,c_T^{}\,$ and $\,c_L^{}\,$ terms are not sizable,
after imposing the experimental data
$\,\MW = 80.385 \!\times\! (1 \pm 1.87\!\times\! 10^{-4})$\,GeV
(which is much more precise than the Higgs observables
to be measured at the future Higgs factory), we can estimate the limit
on the new physics scale to be
$\,\Lambda/\!\sqrt{|c_T^{}|} > 14.9\,(7.59)\,$TeV at $1\sigma$\,(95\%\,C.L.).
This demonstrates significant improvement of the new physics reach
from the precision measurements of the EWPO.
After further including the CEPC measurement of $\sigma(Zh)$,
we find the improved limit, $\,\Lambda/\!\sqrt{|c_T^{}|} > 10.6\,$TeV at 95\%\,C.L.,
as shown in \gtab{tab:steps}.

\begin{table}[t]
\centering
\caption{Impacts of adding the current electroweak precision observables
         ($\alpha, \GF, \MZ, \MW$) \cite{PDG14}
         on probing the new physics scales $\,\Lambda/\!\sqrt{|c_j|}$ (in TeV) at 95\%\,C.L.
         The limits in the first row are
         obtained from $\sigma(Zh)$ to be measured at the CEPC\,\cite{CEPC} only. The limits in
         the second row are given by combining with the current $\MW$ measurement
         plus $\sigma(Zh)$.\, Finally, the third row presents the limits by including
         the current measurements of $(\alpha, \GF, \MZ)$ altogether.
         In the first two rows, $(\alpha, \GF, \MZ)$ are fixed to their experimental central values
         as in the $Z$-scheme, while the third row adopts the scheme-independent approach
         by allowing all electroweak parameters to freely vary in each fit.
         We label the entries of most significant improvements in red color with an underscore.}
\vspace*{3mm}
\begin{tabular}{ccccccccccc}
\hline\hline
  $\mathcal O_H^{}$ & $\mathcal O_T^{}$
& $\mathcal O_{WW}^{}$ & $\mathcal O_{BB}^{}$
& $\mathcal O_{WB}^{}$ & $\mathcal O_{HW}^{}$
& $\mathcal O_{HB}^{}$ & $\mathcal O^{(3)}_{LL}$
& $\mathcal O^{(3)}_L$ & $\mathcal O_L^{}$ & $\mathcal O_R^{}$
\\
\hline
 2.48 &       2.01  & 4.83 & 0.89  &       1.86  & 2.09 & 0.567 &       5.38  & 11.6 & 10.2 & 8.78 \\
 2.48 & \ured{10.6} & 4.83 & 0.89  & \ured{5.16} & 2.09 & 0.567 & \ured{8.22} & 12.1 & 10.2 & 8.78 \\
 2.48 &       10.6  & 4.83 & 0.875 &       5.12  & 2.09 & 0.567 &       8.15  & 12.1 & 10.2 & 8.78 \\
\hline\hline
\end{tabular}
\label{tab:steps}
\label{tab:4}
\end{table}

In \gtab{tab:steps}, the first two rows are essentially $Z$-scheme approach
with $(\alpha, \GF, \MZ)$ fixed. Here we see whether including the
current $\MW$ measurement or not leads to significant difference.
The change appears in the probed new physics scales of the four operators
$\mathcal O_T^{}$, $\mathcal O_{WB}^{}$, $\mathcal O^{(3)}_{LL}$,
and $\mathcal O^{(3)}_{L}$,\,
which are involved in the $Z$-scheme correction \geqn{eq:del-MW-Zscheme}.
For them, the most significant changes come from
$\mathcal O_T^{}$ and $\mathcal O_{WB}^{}$,
since the reaches of the corresponding new physics scales
are enhanced by about a factor of $5$ and $3$, respectively.
It shows that for $\mathcal O_T^{}$, the probe of its new physics scale
is enhanced from $2.01\,\mbox{TeV}$
to $10.6\,\mbox{TeV}$ once $\MW$ measurement is included.
Setting the most precisely measured observables
$(\alpha, \GF, \MZ)$ be their experimental central values
is equivalent to fixing the electroweak observables. This justifies
the $Z$-scheme approach when the precisions of $(\alpha, \GF,\, \MZ)$ are much higher than the
others. In \gsec{sec:ZWmass}, we will further analyze how the situation changes
when the precisions of $\MZ$ and $\MW$ measurements become comparable with each other.

\vspace*{2mm}
\subsection{Enhanced Sensitivity from CEPC Measurements of ${W/Z}$ Masses}
\label{sec:ZWmass}
\label{sec:4.5}
\vspace*{1.5mm}

Lepton colliders such as the CEPC, FCC-ee and ILC
can also make $Z$-pole measurements, which are necessary for
calibrations at the initial stage of running the machine.
To make full use of the $Z$-pole running, we can utilize
the $Z$-pole data to further enhance the indirect probe of new physics scales.
The most significant improvements include the weak boson masses $\MZ$ and $\MW$,
as shown in \gtab{tab:ZWmass} for the CEPC.

\begin{table}[b]
\centering
\caption{Projected precisions (68\%\,C.L.) of $Z$ and $W$ mass measurements
         at the CEPC \cite{CEPC,Zpole}.}
\vspace*{3mm}
\begin{tabular}{c|c|c}
\hline \hline
~Observables~ & Relative Error & ~Absolute Error~ \\
\hline \hline
$M_Z^{}$ & $(0.55  - 1.1) \!\times\! 10^{-5}$ & $(0.5 - 1)$\,MeV \\
$M_W^{}$ & $(3.7 - 6.2) \!\times\! 10^{-5}$ & $(3 - 5)$\,MeV \\
\hline\hline
\end{tabular}
\label{tab:ZWmass}
\label{tab:5}
\end{table}

In comparison with the existing precision data shown in the first block of \gtab{tab:inputs},
we see that the uncertainties of $\,\MZ\,$ and $\,\MW\,$ can be further improved by a
factor of $\,2-4\,$ and $\,3-5$,\, respectively. Since the constraints from current precision
measurements are already rather sensitive, we can expect more significant enhancements
by imposing the CEPC measurements. A rough estimate leads us to expect that the sensitivity
to new physics scales could be doubled for operators $\mathcal O_T^{}$ and $\mathcal O_L^{}$,\,
reaching about $20\,\mbox{TeV}$.

In \gtab{tab:ZWmass-effects}, we quantitatively analyze
the impacts of imposing the $Z$-pole measurements of $\MZ$ and $\MW$
at the CEPC. In the following analysis, we implement the relative errors
$\,8.25 \!\times\! 10^{-6}\,$ for $\MZ$ and
$\,3.7 \!\times\! 10^{-5}\,$ for $\MW$ as an illustration.
Here, we see that the relative errors of $\MZ$ and
$M_W$ become comparable with each other.
Including $\MZ$ alone makes no significant improvement. As we demonstrated in
\gtab{tab:steps} and the related discussions, the effect of inputting the precision
data $\MZ$ is to fix one of the three electroweak parameters.
Adding a better measurement of $\MZ$ would not change this picture, except to further enhance it.
On the other hand, imposing the CEPC measurement of $\MW$  alone can significantly
improve the reach of new physics scales.
This increases the sensitivities to the scales of $\mathcal O_T^{}$,
$\mathcal O_{WB}^{}$, $\mathcal O_L^{}$, and $\mathcal O^{(3)}_{LL}$
by about a factor of two, as shown in the third row of Table\,\ref{tab:6}.
This result is consistent with what we have observed in \gtab{tab:steps}.
A new point is that further imposing the CEPC measurement of $\MZ$, after imposing
$\MW$,\, can introduce extra improvement, although adding the CEPC measurement of $\MZ$ alone
cannot. It demonstrates the fact that when the precisions of $\MZ$ and $\MW$ are comparable
with each other, it is no longer appropriate to just pick up the three observables
to fix the three electroweak variables. In other words, $Z$-scheme is a good approximation
when the relative errors of $(\alpha, \GF, \MZ)$ are all much smaller than the others.
This appears no longer the case at future lepton colliders.
Here, we use the projected CEPC sensitivities to $\MZ$ and $\MW$  \cite{CEPC,Zpole}
as an illustration, and we have demonstrated that the present scheme-independent approach is
a more general-purpose method. In the conventional $Z$-scheme, $\MZ$ is commonly fixed
to the experimental central value, so that the above improvement is impossible.

\begin{table}[t]
\setlength{\tabcolsep}{0.8mm}
\centering
\caption{Impacts of the projected $\MZ$ and $\MW$ measurements at CEPC \cite{CEPC,Zpole} on
         the reach of new physics scale $\Lambda/\!\sqrt{|c_j|}$ (in TeV) at 95\%\,C.L.
         The Higgs observables (including $\sigma(\nu \bar \nu h)$ at 350\,GeV)
         and the existing electroweak precision
         observables (Table\,\ref{tab:inputs}) are always included in each row.
         The differences among the four rows arise from whether taking into account
         the measurements of $\MZ$ and $\MW$ (Table\,\ref{tab:ZWmass}) or not.
         The second (third) row contains the measurement of $\MZ$\,($\MW$) alone,
         while the first (last) row contains none (both) of them.
         We mark the entries of the most significant improvements from
         $\MZ$ and/or $\MW$ measurements in red color with an underscore.}
\vspace*{3mm}
\begin{tabular}{cccccccccccccccc}
\hline \hline
$\mathcal O_H^{}$ & $\mathcal O_T^{}$ & $\mathcal O_{WW}^{}$
& $\mathcal O_{BB}^{}$ & $\mathcal O_{WB}^{}$ & $\mathcal O_{HW}^{}$
& $\mathcal O_{HB}^{}$ & $\mathcal O^{(3)}_{LL}$ & $\mathcal O^{(3)}_L$
& $\mathcal O_L^{}$ & $\mathcal O_R^{}$ & $\mathcal O^{(3)}_{L,q}$
& $\mathcal O_{L,q}^{}$ & $\mathcal O_{R,u}^{}$ & $\mathcal O_{R,d}^{}$
& $\mathcal O_g^{}$ \\
\hline
2.74 & 10.6  & 6.38 & 5.78 & 6.53  & 2.16 & 0.604 & 8.58  & 12.1 & 10.2 & 8.78 & 2.06 & 0.568 & 0.393 & 0.339 & 43.8
\\
2.74 & \ured{10.7} & 6.38 & 5.78 & \ured{6.54} & 2.16 & 0.604 & \ured{8.62} &  12.1
& 10.2 & 8.78 & 2.06 & 0.568 & 0.393 & 0.339 & 43.8
\\
2.74 & \ured{21.0} & 6.38 & 5.78 & \ured{10.4} & 2.16 & 0.604 &
\ured{15.5} & \ured{16.4} & 10.2 & 8.78 & 2.06 & 0.568 & 0.393 & 0.339 & 43.8
\\
2.74 & \ured{23.7} & 6.38 & 5.78 & \ured{11.6} & 2.16 & 0.604 & \ured{17.4} & \ured{18.1}
& 10.2 & 8.78 & 2.06 & 0.568 & 0.393 & 0.339 & 43.8
\\
\hline\hline
\end{tabular}
\label{tab:ZWmass-effects}
\label{tab:6}
\vspace*{2mm}
\end{table}

\vspace*{2mm}
\subsection{Enhancement from $Z$-Pole Observables at CEPC}
\label{sec:Zpole}
\label{sec:4.6}
\vspace*{1.5mm}

In addition to the mass measurements of $W$ and $Z$,
CEPC can also measure the $Z$ boson lineshape
at the $Z$-pole, $\sqrt s = \MZ$\,.\,
Currently, there are six observables that have been simulated
at CEPC \cite{CEPC,Zpole}. For
convenience, we summarize them in \gtab{tab:Zpole}, in the order of their relative precisions.

\begin{table}[t]
\centering
\caption{Projected precisions (68\%\,C.L.) of $Z$-pole measurements at the CEPC \cite{CEPC,Zpole}.}
\vspace*{3mm}
\begin{tabular}{c|c}
\hline \hline
~Observables~ & ~Relative Error~ \\
\hline \hline
$N_\nu$           & $1.8 \times 10^{-3}$ \\
$A_{FB}(b)$       & $1.5 \times 10^{-3}$ \\
$R_b^{}$             & $8 \times 10^{-4}$ \\
$R_\mu^{}$           & $5 \times 10^{-4}$ \\
$R_\tau^{}$          & $5 \times 10^{-4}$ \\
$\sin^2 \theta_W^{}$ & $1 \times 10^{-4}$ \\
\hline\hline
\end{tabular}
\label{tab:Zpole}
\label{tab:7}
\end{table}

In comparison with the existing measurements of LEP \cite{PDG14},
CEPC can improve the accuracy by at least one order of magnitude.
The relative errors of the projected CEPC measurements range
from $\,1.8 \!\times\! 10^{-3}\,$ to $\,10^{-4}\,$ as shown in \gtab{tab:Zpole}.
Although these relative errors appear larger than those of the mass measurements
for $Z$ and $W$ bosons,
they are still much smaller than the Higgs observables listed in \gtab{tab:inputs}.
The most sensitive Higgs observable at the CEPC is the production cross section
$\,\sigma(Zh)\,$,\,  which can be measured to the precision of $\,0.51\%\,$.\,
We can expect a much more improved constraint on the new physics scales
by using the $Z$-pole observables.

For this analysis, we derive the linearly expanded
expressions for the new physics contributions to the observables
shown in \gtab{tab:Zpole}.  We use the analytical formulae of
these observables given in \cite{LEP1}.
The new physics enters these observables through the parameter shifts of
the involved vertices between the $Z$ boson and fermions. Since the deviations
from the SM predictions should be reasonably small, we can expand the parameter shifts
up to the linear order. For convenience, we present the expanded expressions as follows,
\begin{subequations}
\begin{eqnarray}
  \frac {\delta \widetilde {N_\nu}}{N_\nu}
& = &
  2 \frac{\,\delta\GF\,}{\GF}
+ 5 \frac{\,\delta\MZ\,}{\MZ}
- 0.0908 \frac{c_T^{}}{\,\LambdaTeV^2\,}
\nonumber
\\
&&
+ 0.103 \frac{\,c_{WW}^{}\,}{\,\LambdaTeV^2\,}
+ 0.00747 \frac{\,c_{BB}^{}\,}{\,\LambdaTeV^2\,}
+ 0.0277 \frac{\,c_{WB}^{}\,}{\,\LambdaTeV^2\,}
+ 0.121 \frac{c^{(3)}_{LL}}{\,\LambdaTeV^2\,}
- 0.121 \frac{\,c_L^{}\,}{\,\LambdaTeV^2\,} ,~~~~~
\qquad
\\
  \frac {\delta \widetilde {A_{FB}(b)}}{A_{FB}(b)}
& = &
  7.5 \frac {\delta \GF}{\GF}
+ 15 \frac {\delta\MZ}{\MZ}
- 7.5 \frac {\delta\alpha} \alpha
+ 0.391 \frac {c_{WW}^{}}{\,\LambdaTeV^2\,}
- 0.0488 \frac {c_{BB}^{}}{\,\LambdaTeV^2\,}
- 0.038 \frac {c_{WB}^{}}{\,\LambdaTeV^2\,}
\nonumber
\\
&&
+ 0.324 \frac {c^{(3)}_L}{\,\LambdaTeV^2\,}
+ 0.324 \frac {c_L^{}}{\,\LambdaTeV^2\,}
+ 0.44 \frac {c_R^{}}{\,\LambdaTeV^2\,}
\nonumber
\\
&&
- 0.00766 \frac {c^{(3)}_{L,q}}{\,\LambdaTeV^2\,}
+ 0.00766 \frac {c_{L,q}^{}}{\,\LambdaTeV^2\,}
+ 0.0465 \frac {c_{R,d}^{}}{\,\LambdaTeV^2\,} ,
\\
\frac {\delta \widetilde {R_b^{}}}{R_b^{}}
& = &
- 0.0658 \frac {\delta \GF}{\GF}
- 0.117 \frac {\delta \MZ}{\MZ}
+ 0.0658 \frac {\delta \alpha} \alpha
- 0.000451 \frac {c_T^{}}{\,\LambdaTeV^2\,}
\nonumber
\\
&&
+ 0.00268 \frac {c_{WW}^{}}{\,\LambdaTeV^2\,}
+ 0.000872 \frac {c_{BB}^{}}{\,\LambdaTeV^2\,}
+ 0.00198 \frac {c_{WB}^{}}{\,\LambdaTeV^2\,}
\nonumber
\\
&&
- 0.0976 \frac {c^{(3)}_{L,q}}{\,\LambdaTeV^2\,}
+ 0.0976 \frac {c_{L,q}^{}}{\,\LambdaTeV^2\,}
- 0.0198 \frac {c_{R,u}^{}}{\,\LambdaTeV^2\,}
- 0.00703 \frac {c_{R,d}^{}}{\,\LambdaTeV^2\,} ,
\\
\frac {\delta \widetilde {R_\mu}^{}}{R_\mu^{}}
& = &
  0.0923 \frac {\delta \GF}{\GF}
+ 0.189 \frac {\delta \MZ}{\MZ}
- 0.0923 \frac {\delta \alpha} \alpha
- 0.000138 \frac {c_T^{}}{\,\LambdaTeV^2\,}
\nonumber
\\
&&
+ 0.0253 \frac {c_{WW}^{}}{\,\LambdaTeV^2\,}
+ 0.000887 \frac {c_{BB}^{}}{\,\LambdaTeV^2\,}
+ 0.00506 \frac {c_{WB}^{}}{\,\LambdaTeV^2\,}
\nonumber
\\
&&
- 0.136 \frac {c^{(3)}_L}{\,\LambdaTeV^2\,}
- 0.136 \frac {c_L^{}}{\,\LambdaTeV^2\,}
+ 0.1 \frac {c_R^{}}{\,\LambdaTeV^2\,}
\nonumber
\\
&&
- 0.0398 \frac {c^{(3)}_{L,q}}{\,\LambdaTeV^2\,}
+ 0.0398 \frac {c_{L,q}^{}}{\,\LambdaTeV^2\,}
+ 0.0198 \frac {c_{R,u}^{}}{\,\LambdaTeV^2\,}
- 0.0146 \frac {c_{R,d}^{}}{\,\LambdaTeV^2\,} ,
\\
\frac {\delta \widetilde {R_\tau}^{}}{R_\tau^{}}
& = &
  0.0915 \frac {\delta \GF}{\GF}
+ 0.183 \frac {\delta \MZ}{\MZ}
- 0.0915 \frac {\delta \alpha} \alpha
+ 0.0252 \frac {c_{WW}^{}}{\,\LambdaTeV^2\,}
\nonumber
\\
&&
+ 0.000886 \frac {c_{BB}^{}}{\LambdaTeV^2}
+ 0.00504 \frac {c_{WB}^{}}{\LambdaTeV^2}
- 0.136 \frac {c^{(3)}_L}{\LambdaTeV^2}
- 0.136 \frac {c_L^{}}{\LambdaTeV^2}
+ 0.1 \frac {c_R^{}}{\LambdaTeV^2}
\nonumber
\\
&&
- 0.0398 \frac {c^{(3)}_{L,q}}{\LambdaTeV^2}
+ 0.0398 \frac {c_{L,q}^{}}{\LambdaTeV^2}
+ 0.0198 \frac {c_{R,u}^{}}{\LambdaTeV^2}
- 0.0146 \frac {c_{R,d}^{}}{\LambdaTeV^2} ,
\\
  \frac {\delta \widetilde {\sin^2 \theta_W^{}}}{\sin^2 \theta_W^{}}
& = &
- 1.37 \frac {\delta \GF}{\GF}
- 2.74 \frac {\delta \MZ}{\MZ}
+ 1.37 \frac {\delta \alpha} \alpha
- 0.0692 \frac {c_{WW}^{}}{\,\LambdaTeV^2\,}
+ 0.00907 \frac {c_{BB}^{}}{\,\LambdaTeV^2\,}
\nonumber
\\
&&
+ 0.00753 \frac {c_{WB}^{}}{\,\LambdaTeV^2\,}
- 0.0605 \frac {c^{(3)}_L}{\,\LambdaTeV^2\,}
- 0.0605 \frac {c_L^{}}{\,\LambdaTeV^2\,}
- 0.0821 \frac {c_R^{}}{\,\LambdaTeV^2\,} .
\end{eqnarray}
\end{subequations}

\begin{table}[t]
\setlength{\tabcolsep}{0.8mm}
\centering
\caption{Impacts of the projected $Z$-pole measurements at the CEPC \cite{CEPC,Zpole}
         on the reach of new physics scale $\Lambda/\sqrt{|c_j|}$ (in TeV) at 95\%\,C.L.
         For comparison, the first row of this table repeats the last row of
         Table\,\ref{tab:6}, as our starting point of this table.
         For the $(n+1)$-th row, the first $n$ observables in Table\,\ref{tab:Zpole} are
         taken into account. In addition, the estimated $\MZ$ and $\MW$ measurements at the CEPC
         in Table\,\ref{tab:ZWmass}, the Higgs observables (HO), and the existing electroweak precision
         observables (EWPO) in Table\,\ref{tab:inputs} are always included for each row.
         The entries with major enhancements of the new physics scale limit are marked
         in red color with an underscore.}
\vspace*{3mm}
\begin{tabular}{cccccccccccccccc}
\hline \hline
 $\mathcal O_H^{}$ & $\mathcal O_T^{}$ & $\mathcal O_{WW}^{}$
 & $\mathcal O_{BB}^{}$ & $\mathcal O_{WB}^{}$ & $\mathcal O_{HW}^{}$
 & $\mathcal O_{HB}^{}$ & $\mathcal O^{(3)}_{LL}$ & $\mathcal O^{(3)}_L$
 & $\mathcal O_L^{}$ & $\mathcal O_R^{}$ & $\mathcal O^{(3)}_{L,q}$
 & $\mathcal O_{L,q}^{}$ & $\mathcal O_{R,u}^{}$
 & $\mathcal O_{R,d}^{}$ & $\mathcal O_g^{}$ \\
\hline
 2.74 & 23.7 & 6.38 & 5.78 & 11.6 & 2.16 & 0.604 & 17.4 & 18.1 & 10.2 & 8.78 & 2.06 & 0.568 & 0.393 & 0.339 & 43.8 \\
\hline
 2.74 & 23.7 & 6.38 & 5.78 & 11.6 & 2.16 & 0.604 & \ured{17.5} & \ured{18.3} & \ured{10.5} & 8.78 & 2.06 & 0.568 & 0.393 & 0.339 & 43.8 \\
 2.74 & 24.0 & \ured{8.32} & 5.80 & \ured{12.2} & 2.16 & 0.604 & \ured{20.7} & \ured{23.0} & \ured{12.5} & \ured{13.0} & 2.23 & \ured{1.62} & 0.393 & \ured{3.97} & 43.8 \\
 2.74 & 24.0 & 8.33 & 5.80 & 12.2 & 2.16 & 0.604 & 20.7 & 23.0 & 12.5 & 13.0 & \ured{7.90} & \ured{7.89} & \ured{3.55} & 4.05 & 43.8 \\
 2.74 & 24.0 & 8.54 & 5.80 & 12.2 & 2.16 & 0.604 & 20.7 & 23.4 & \ured{14.4} & \ured{14.0} & 8.63 & 8.62 & 4.88 & \ured{4.71} & 43.8 \\
 2.74 & 24.0 & 8.75 & 5.81 & 12.3 & 2.16 & 0.604 & 20.7 & 23.7 & 15.8 & 14.9 & \ured{9.21} & \ured{9.21} & 5.59 & 5.17 & 43.8 \\
 2.74 & \ured{26.3} & \ured{12.6} & \ured{5.93} & \ured{15.3} & 2.16 & 0.604 & \ured{30.2} & \ured{35.2} & \ured{19.8} & \ured{21.6} & 9.21 & 9.21 & 5.59 & 5.17 & 43.8 \\
\hline\hline
\end{tabular}
\label{tab:Zpole-effects}
\label{tab:8}
\vspace*{7mm}
\end{table}

We see that these observables involve almost
all dimension-6 operators in \gtab{tab:O}, except the pure-Higgs operator
$\,\mathcal O_H^{}\,$ and
the gluon operator $\,\mathcal O_g^{}\,$.
The bosonic operators $\,\mathcal O_T^{}\,$,\, $\mathcal O_{WW}^{}$,\,
$\mathcal O_{BB}^{}$,\, and $\,\mathcal O_{WB}^{}$\,
can enter through the field redefinitions and mass shifts.
Only the operators $\mathcal O_{HB}^{}$ and $\mathcal O_{HW}^{}$ are not involved.

\begin{figure}[t]
\centering
\includegraphics[height=9cm]{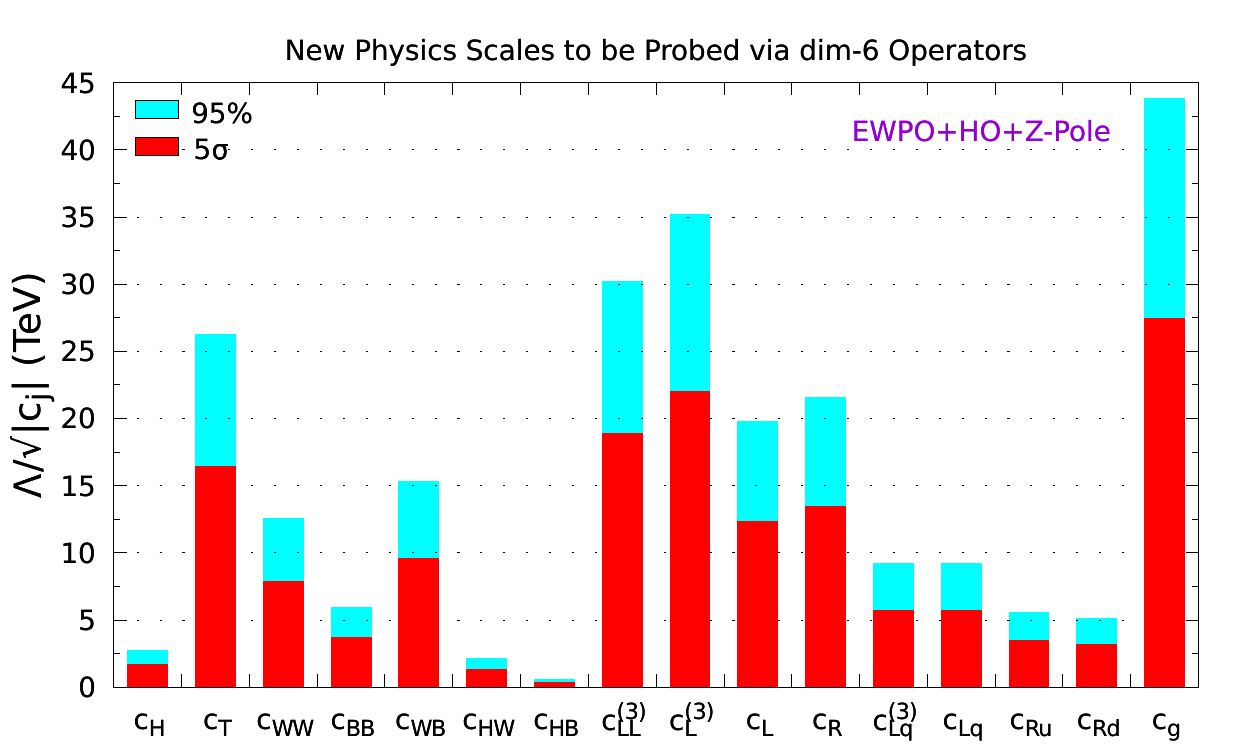}
\caption{The 95\% exclusion (blue) and $5\sigma$ discovery (red) sensitivities
to the new physics scales $\,\Lambda/\!\sqrt{|c_j|}\,$
by combining the current electroweak precision measurements ($\alpha,\,\GF,\,\MZ,\,\MW$) \cite{PDG14}
with the future Higgs observables at the Higgs factory CEPC (Table\,\ref{tab:inputs}) and $Z$-pole
measurements (Table\,\ref{tab:ZWmass}) under a projected luminosity of \,5\,ab$^{-1}$ \cite{CEPC}.}
\label{fig:4}
\end{figure}

In \gtab{tab:Zpole-effects}, we present the sensitivity reaches
by including the $Z$-pole observables summarized
in \gtab{tab:Zpole}. The $n$-th row corresponds to the constraint from the
$(Z,W)$ mass measurements, the Higgs observables, and the existing EWPO,
plus the first $n$ observables in \gtab{tab:Zpole}.
The difference between the $(n)$-th and $(n\!+\!1)$-th rows represents
the effect of the $n$-th $Z$-pole observable in \gtab{tab:7}.
It is striking to see that including the CEPC $Z$-pole measurements can
further probe the new physics scale up to
$\,35\,\mbox{TeV}$\, for $\mathcal O^{(3)}_L$.\,
This is another factor-2 enhancement over that of
only including $(Z,W)$ mass measurements in \gtab{tab:ZWmass-effects}.
The relative enhancements to the scales of $\mathcal O_{WW}^{}$, $\mathcal O_R^{}$,
$\mathcal O^{(3)}_{L,q}$, $\mathcal O_{L,q}^{}$, $\mathcal O_{R,u}^{}$,
and $\mathcal O_{R,d}^{}$ are even larger,
while operators $\mathcal O_{WB}^{}$, $\mathcal O^{(3)}_{LL}$, and $\mathcal O_L^{}$
also receive significantly enhanced constraints.
In contrast, the operator $\mathcal O_{BB}^{}$ is not significantly improved since
its contribution to $Z$-pole observables is highly suppressed.
We present the final results in \gfig{fig:4}.

\gfig{fig:4} demonstrates that the $Z$-pole measurements are even more sensitive
than the Higgs observables for indirectly constraining the new physics scales
of effective dimension-6 operators. This is mainly because of the huge
event number that can be produced at the $Z$-pole resonance.
We see that running the future $e^+ e^-$ collider at $Z$-pole
is beyond the technical purpose of the machine calibration.
Our study shows that it is worth of running the collider at $Z$-pole for a longer time.
Or, after running the Higgs factory at Higgsstrahlung energy ($240-250$\,GeV),
it is invaluable to return to the $Z$-pole running for a period and thus ensure
the no-lose probe of new physics.

\vspace*{3mm}
\section{Higgs Coupling Precision Tests at CEPC and Probing Dimension-6 Yukawa-type Operators}
\label{sec:5}
\label{app:cepc}
\label{app:C}
\vspace*{2mm}

In this section, we study the CEPC sensitivities to the SM-type Higgs couplings,
and then apply these limits to study the probe of Yukawa-type dimension-6 operators
(cf.\ \gtab{tab:1}).
In \gsec{sec:5.1}, we first apply our analytical linear $\chi^2$ fitting method in \gapp{sec:chi2}
to study the sensitivity probe of the SM-type Higgs couplings at the CEPC.
Then, based upon these, we will analyze
the CEPC reach of new physics scales associated with
the Yukawa-type dimension-6 operators in \gsec{sec:5.2}.

\vspace*{2mm}
\subsection{Higgs Coupling Precision Tests at CEPC}
\label{sec:5.1}
\vspace*{1.5mm}

For an illustration, we apply the analytical linear $\chi^2$ fitting method described in \gapp{sec:chi2}
to extract the projected precisions of the CEPC Higgs measurements for probing the SM-type Higgs couplings.
The Higgs couplings to other SM particles may be defined relative
to their SM values by rescaling,\, $g_{hii}^{} / g^{\text{sm}}_{hii} \equiv \kappa_i^{}\,$,
where the possible deviation $\,\kappa_i^{}-1\,$ denotes the anomalous Higgs couplings.
Such deviations $\,\kappa_i^{}-1\,$ can arise from the dimension-6 operators
shown in Table\,\ref{tab:1}.  The anomalous Higgs couplings $\,\kappa_i^{}\neq 1\,$ will modify
the Higgs observables in \gtab{tab:inputs} and thus receive constraints by the CEPC measurements.

The cross sections of Higgsstrahlung and $WW$ fusion processes are
scaled by the Higgs couplings with $Z$ and $W$ gauge bosons as
$\,\delta \sigma(Zh) / \sigma(Zh) \simeq 2 \delta \kappa_Z^{}\,$ and
$\,\delta \sigma(\nu \nu h) / \sigma(\nu \nu h) \simeq 2 \delta \kappa_W^{}$.\,
On the other hand, each partial decay width of $\,h\to ii\,$ scales as,
$\Gamma_{hii}^{} / \Gamma^{\text{sm}}_{hii} = \kappa^2_i$.\,
For the exotic decay channels which are not present in the SM, such as the invisible decays,
we can parametrize its contribution as a fraction of the total SM Higgs decay width,
$\Gamma_{\inv}/\Gamma^{\sm}_{\textrm{tot}} = \BR(\inv) \equiv \delta \kappa_{\inv}$,
which is relatively small deviation in principle.
Each branching fraction $\,\BR_i^{}\,$ is a ratio between the individual decay width and total width,
and is thus a function of all scaling factors $\{\kappa_i^{}\}$,
Since so far the SM fits LHC data quite well and the CEPC measurements can be rather precise,
we expect that the relative deviations from the SM are significantly below one,
$\,|\kappa_i^{}-1| \ll 1$\,.\,
We thus define,
$\, \kappa_i^{} \equiv 1 + \delta \kappa_i^{}
$,\,
with $\,|\delta \kappa_i^{}| \ll 1$.\,
Thus, we may expand the branching fractions up to the linear order of
$\,\delta \kappa_i^{}\,$,
\begin{equation}
  \BR^\th_i \,\simeq\,  \BR^{\th,0}_i
\Big( 1 + \sum_j A_{ij}^{} \delta \kappa_j^{} \Big) ,
\qquad~~
  \BR^\th_{\inv} \,\simeq\,
  \delta \kappa_{\inv}^{} ,
\label{eq:BR-expanded}
\end{equation}
where $\,\BR^{\th,0}_i = \BR^{\sm}_i\,$ is the SM prediction, and the coefficient matrix $\,A$\, is,
\begin{equation}
A_{ij}^{} =
  2 ( \delta_{ij}^{} - \BR^{\sm}_j) ,
\qquad
  A_{i, \inv}^{} =  - 1 ,
\qquad
  A_{\inv, i}^{} =   0 ,
\qquad
  A_{\inv, \inv}^{} =  1 .
\label{eq:A}
\end{equation}
Note that different branching fractions are correlated with coefficient proportional
to the corresponding SM values, as shown in \geqn{eq:Br-i}. For
the branching fraction $\BR_i^{}$, the contribution due to its own channel is modulated
by $\,1 - \Br^{\text{sm}}_i\,$,\, while the effect from other channels by the corresponding
$\,\BR^{\text{sm}}_i$.\, Larger branching fraction means the channel has smaller effect on
its own, but larger on the others.

Applying our analytical $\chi^2$ fitting method (\gapp{sec:chi2}) together with
the relative uncertainties of Higgs production cross sections and branching fractions
from \gtab{tab:inputs}, we extract the sensitivities of CEPC measurements
to the SM Higgs couplings as shown in \gtab{tab:precision2} with two different
fits in the second and third columns. The first is a 9+1 parameter fit, including
9 parameters for decay branching fractions and 1 for total decay width. All the
anomalous Higgs couplings have precisions at $1\%$ level, except that
$\,\kappa_\gamma^{}\,$ and $\,\kappa_\mu^{}\,$ have larger uncertainties.
This is because the branching fractions, $\BR(\gamma \gamma)$
and $\BR(\mu \bar \mu)$, are too small according to the SM predictions \cite{hdecay}.
As shown in \gtab{tab:inputs}, their values are well below 1\%. Since roughly 1 million
Higgs particles can be produced at CEPC \cite{CEPC},
measuring the decays into photon or muon can collect less than $10^4$ events.
The statistical fluctuation is thus larger than 1\%.\,
A realistic estimate gives \,9\% and 17\%,\, respectively, including both statistical
and systematic uncertainties. On the contrary, the $ZZh$ Higgs coupling $\,\kappa_Z^{}\,$
has a precision much better than $1\%$,\, due to the direct measurement of
the Higgsstrahlung production cross section $\sigma(Zh)$. This inclusive
production rate has larger event rate than any individual decay channel.
Without $\sigma(Zh)$, the precision on $\kappa_Z^{}$ is also at percentage level.
The same thing applies to $\kappa_W^{}$,\, which can be constrained by
the $WW$ fusion production rate $\sigma(\nu \nu h)$, leading to roughly
a factor of $\sqrt 2$\, improvement.

\begin{table}[t]
\centering
\caption{Projected precisions of measuring Higgs couplings (68\%\,C.L.)
         at the CEPC\,(250GeV, 5ab$^{-1}$) from our fit, in comparison with
         the LHC (14TeV, 300fb$^{-1}$), HL-LHC (14TeV, 3ab$^{-1}$) and
         ILC (250GeV, 250fb$^{-1}$)+(500GeV, 500fb$^{-1}$) \cite{Peskin}.}
\vspace*{3mm}
\begin{tabular}{c|cc|cc|cc}
\hline \hline
 \multirow{2}{*}{Precision (\%)} & \multicolumn{2}{c|}{CEPC} & \multirow{2}{*}{LHC} & \multirow{2}{*}{HL-LHC} & \multirow{2}{*}{ILC-250} & \multirow{2}{*}{ILC-500} \\
                                & 9+1 fit & 8+1 fit         &                     &                        &                         &  \\
\hline \hline
 $\kappa_Z^{}$      & 0.249 & 0.249 &  8.5 & 6.3 & 0.78 & 0.50 \\
 $\kappa_W^{}$      &  1.20 &  1.20 &  5.4 & 3.3 &  4.6 & 0.46 \\
 $\kappa_\gamma^{}$ &  4.67 &  4.67 &  9.0 & 6.5 & 18.8 & 8.6  \\
 $\kappa_g^{}$      &  1.42 &  1.42 &  6.9 & 4.8 &  6.1 & 2.0  \\
 $\kappa_b^{}$      &  1.27 &  1.27 & 14.9 & 8.5 &  4.7 & 0.97 \\
 $\kappa_c^{}$      &  1.75 &  1.75 & --   & --  &  6.4 & 2.6  \\
 $\kappa_\tau^{}$   &  1.33 &  1.33 &  9.5 & 6.5 &  5.2 & 2.0  \\
 $\kappa_\mu^{}$    &  8.59 &  --   & --   & --  &  --  & --   \\
 $\BR(\inv)$        & 0.134 & 0.134 &  8.0 & 4.0 & 0.54 & 0.52 \\
 $\Gamma_h^{}$      & 2.6   &  2.6  & --   & --   &  -- & --
 \\[1mm]
\hline \hline
\end{tabular}
\label{tab:precision2}
\label{tab:9}
\end{table}

\begin{figure}[t]
\centering
\includegraphics[width=14.5cm,height=9cm]{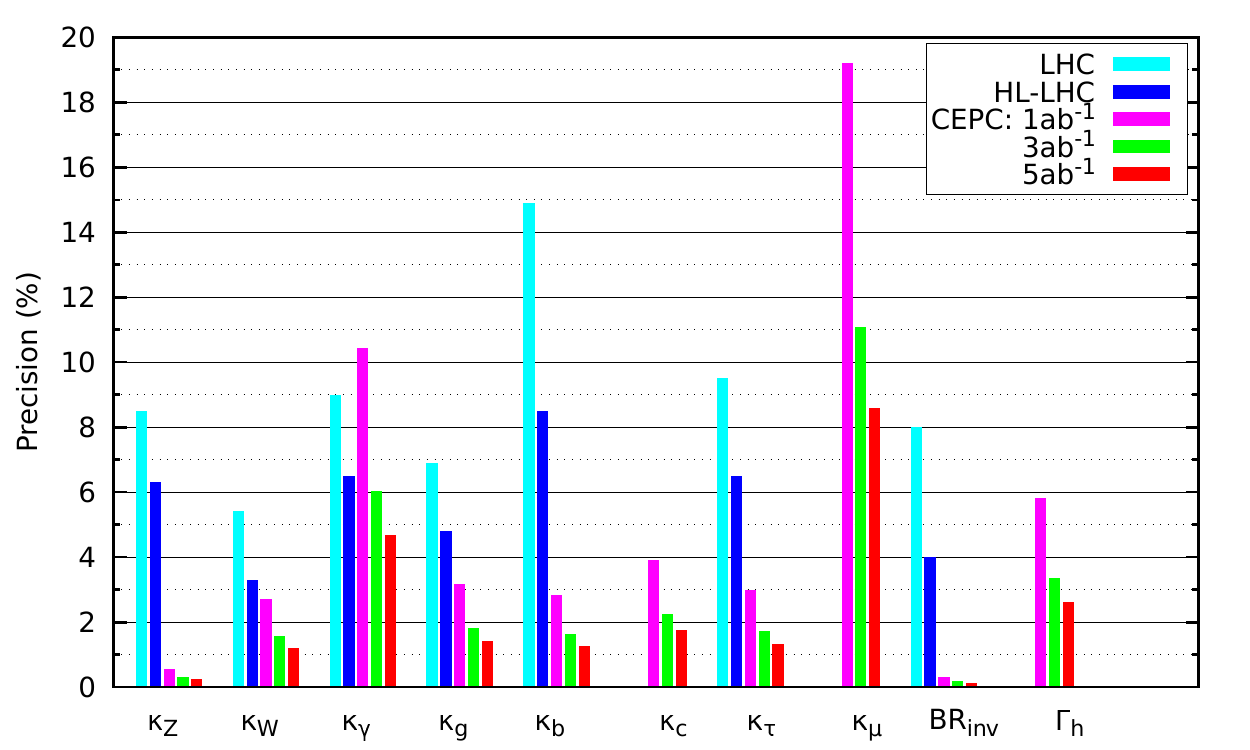}
\vspace*{-2mm}
\caption{Precisions (68\%\,C.L.) of the CEPC\,(250\,GeV) for measuring the
Higgs gauge couplings and Yukawa couplings from our 9+1 parameter fit,
with an integrated luminosity of \,(1,\,3,\,5)\,ab$^{-1}$,\, respectively.
These are compared to the precisions of the LHC (14\,TeV, 300\,fb$^{-1}$) and
HL-LHC (14\,TeV, 3\,ab$^{-1}$) \cite{Peskin}.}
\label{fig:fit91}
\label{fig:5}
\vspace*{2mm}
\end{figure}

For comparison, we further present an 8+1 parameter fit with $\BR(\mu \bar \mu)$ and
$\kappa_\mu^{}$ removed in the third column of \gtab{tab:precision2}. We note that
the precision of measuring other anomalous couplings are not affected at all.
This is because the branching fraction of this channel is very small in the first place.
As explained below Eq.\,\geqn{eq:A}, the correlation
is proportional to the corresponding SM prediction $\,\BR^{\text{sm}}_j$.\,
Hence, it is rather weakly correlated with other channels.
We present the result of these two fits in \gfig{fig:fit91}.

Besides the precision limits on Higgs couplings
at the CEPC\,(250\,GeV, 5\,ab$^{-1}$), we also show
the bounds on Higgs couplings from the LHC\,(14\,TeV, 300\,fb$^{-1}$)
and the HL-LHC\,(14\,TeV, 3\,ab$^{-1}$)
\cite{Peskin}, in \gtab{tab:precision2}, for comparison.
It is clear that the CEPC\,(250\,GeV, 5\,ab$^{-1}$) can significantly improve the
precision of Higgs coupling measurements.
In addition, many decay channels cannot be probed at the LHC.
For instance, the LHC has no sensitivity to the $\,hc\bar{c}\,$ coupling \cite{Peskin},
as well as $h\mu\bar{\mu}$ coupling. But, they can be measured at the CEPC instead.
The total decay width of the SM Higgs with $125\,$GeV mass is about $4$\,MeV,\,
which is far below the LHC sensitivity.
It is hard to make a direct measurement at the LHC without model assumptions.
In \gtab{tab:precision2}, we also show the
projected limits of the ILC\,(250\,GeV, 250\,fb$^{-1}$) and
ILC\,(500\,GeV, 500\,fb$^{-1}$) for comparison \cite{Peskin}.
It shows that CEPC\,(250GeV, 5ab$^{-1}$) can have better sensitivities than the
ILC\,(500GeV, 500fb$^{-1}$),
except for the $hWW$ and $h b \bar b$ couplings.

\vspace*{2mm}
\subsection{Probing Dimension-6 Yukawa-type Operators at CEPC}
\label{sec:5.2}
\vspace*{1.5mm}

The last column of \gtab{tab:O} also presents three Yukawa-type dimension-6
operators $\,\OO_y^f = (\OO_y^u,\, \OO_y^d,\, \OO_y^{\ell})$.\,
These operators will modify the SM Yukawa coupling by a rescaling factor,
\beqa
\label{eq:Yf-d6}
  y_f^{\text{sm}} ~~\longrightarrow~~
  y_f^{} \,=\, y_f^{\text{sm}} + \frac {\,3 c_f^{} v^2\,}{2 \Lambda^2} \,,
\eeqa
and correct the SM fermion mass,
\beqa
\label{eq:mf-d6}
m_f^{\text{sm}} = \frac{y_f^{\textrm{sm}} v}{\sqrt{2}}\,
~~\longrightarrow~~
m_f^{} \,=\, \frac{v}{\sqrt{2}\,}\!
\(\! y_f^{\text{sm}} + \frac{\,c_f^{}v^2\,}{\,2\Lambda^2\,}\!\) \!,
\eeqa
where 
$\,m_f^{}\,$ is the full fermion mass including contributions of dimension-6 operators.
Thus, using \eqref{eq:mf-d6}, we can reexpress $\,y_f^{\text{sm}}\,$ as
\beqa
\label{eq:yf-mf}
y_f^{\textrm{sm}} \,=\, \frac{\,\sqrt{2}m_f^{}\,}{v}
-\frac{\,c_f^{}v^2\,}{\,2\Lambda^2\,}
\,=\, y_f^{\textrm{sm},0} -\frac{\,c_f^{}v^2\,}{\,2\Lambda^2\,} ,
\eeqa
where $\,y_f^{\textrm{sm},0} \equiv \sqrt{2}m_f^{}/v\,$.\,
From Eqs.\,\eqref{eq:Yf-d6} and \eqref{eq:yf-mf},
we compute the coupling ratios up to $\OO(\Lambda^{-2})$,
\beqs
\beqa
\label{eq:kf}
\kappa_f^{} &=& \frac{\,y_f^{}\,}{\,y_f^{\textrm{sm}}\,}
\,\simeq\, 1 + \frac{3c_f^{}v^3}{\,2\sqrt{2}\,m_f^{}\Lambda^2\,} \,,
\\
y_f^{} &=& y_f^{\textrm{sm},0}\( y_f^{}/ y_f^{\textrm{sm},0}\)
\,=\, y_f^{\textrm{sm},0}\tilde{\kappa}_f^{} \,,
\\
\tilde{\kappa}_f^{}
&\equiv& \frac{y_f^{}}{\,y_f^{\textrm{sm},0}\,}
\,=\, \frac{y_f^{\textrm{sm}}}{\,y_f^{\textrm{sm},0}\,} \kappa_f^{}
\,\simeq\,
1 + \frac{c_f^{}v^3}{\,\sqrt{2}\,m_f^{}\Lambda^2\,} \,.
\label{eq:kf-tilde}
\eeqa
\eeqs
We note that it is the effective coupling $\,y_f^{}\,$ that actually enters the physical
observables, and the $\chi^2$ fit we made in \gsec{sec:5.1} (\gtab{tab:9}) is just
a fit of the sensitivity reach on the coupling ratio
$\,\tilde{\kappa}_f^{}\equiv {y_f^{}}/{\,y_f^{\textrm{sm},0}\,}
   \equiv 1+\Delta\tilde\kappa_f^{}\,$,\,
for the case of Higgs Yukawa couplings,
where $\,\Delta\tilde\kappa =\tilde\kappa_f^{}-1\,$
is given by Eq.\,\eqref{eq:kf-tilde} for the contribution of
dimension-6 operator $\,\OO_y^f\,$.\,
This means that each $\,\kappa_f^{}\,$ in \gtab{tab:9} should be replaced
by the current notation $\,\tilde\kappa_f^{}\,$ as we exactly defined in Eq.\,\eqref{eq:kf-tilde}.

\begin{table}[t]
\centering
\caption{Sensitivity reaches (95\%\,C.L.) of the new physics scales of Yukawa-type dimension-6 operators
         at the CEPC\,(250GeV, 5ab$^{-1}$), in comparison with
         the LHC\,(14TeV, 300fb$^{-1}$), HL-LHC\,(14TeV, 3ab$^{-1}$), and
         ILC\,(250GeV, 250fb$^{-1}$)+(500GeV, 500fb$^{-1}$).}
\vspace*{3mm}
\begin{tabular}{c|cc|cc|cc}
\hline \hline
 \multirow{2}{*}{$\Lambda/\!\sqrt{|c_j|} (\mbox{TeV})$}
 & \multicolumn{2}{c|}{CEPC} & \multirow{2}{*}{LHC}
 & \multirow{2}{*}{HL-LHC} & \multirow{2}{*}{ILC-250}
 & \multirow{2}{*}{ILC-500} \\
 & 9+1 fit & 8+1 fit
 & & & &  \\
\hline \hline
 $b$ quark      &  13.2 &  13.2 & 3.87 & 5.12 & 6.89 & 15.2 \\
 $c$ quark      &  24.4 &  24.4 & --   &  --  & 12.8 & 20.0 \\
 $\tau$ lepton  &  15.4 &  15.4 & 5.74 & 6.95 & 7.76 & 12.5 \\
 $\mu$  lepton  &  25.1 &  --   & --   & --   & --   & --
 \\[1mm]
\hline \hline
\end{tabular}
\label{tab:yukawaLike}
\label{tab:10}
\vspace*{3mm}
\end{table}

\begin{figure}[t]
\centering
\includegraphics[width=13cm,height=9cm]{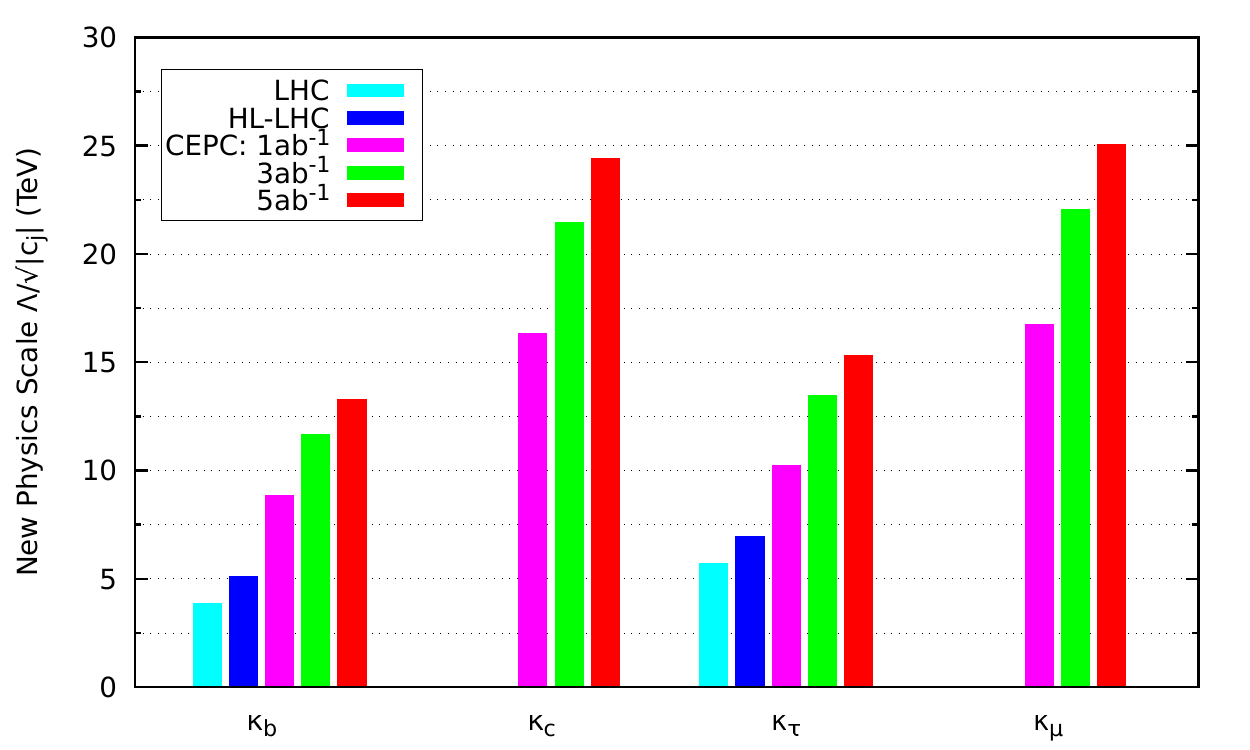}
\vspace*{-2mm}
\caption{Sensitivity reaches (95\%\,C.L.) of the new physics scales $\,\Lambda/\!\sqrt{|c_j|}$\,
         of Yukawa-type dimension-6 operators by the precision Higgs coupling measurements at the
         Higgs factory CEPC\,(250GeV), in comparison with LHC\,(14TeV,\,300fb$^{-1}$) and
         the HL-LHC\,(14TeV,\,3ab$^{-1})$.}
\label{fig:yukawaLike}
\label{fig:6}
\vspace*{3mm}
\end{figure}

Thus, for each given fitted experimental sensitivity
$\,\Delta\tilde\kappa_f^{}\,$ (\gtab{tab:9})
and using Eq.\,\eqref{eq:kf-tilde},
we can derive the following lower bound on the Yukawa-type new physics scale,
\beqa
  \frac \Lambda {\sqrt{|c_f|\,}\,}
 ~\geqq~
  \sqrt{\frac{v^3}{\,\sqrt 2\,m_f^{}\Delta\tilde\kappa_f^{}\,}\,} \,.
\label{eq:yukawaLike}
\label{eq:OY-limit}
\eeqa
In Eq.\,\eqref{eq:yukawaLike}, the Yukawa coupling precision
$\,\Delta\tilde\kappa_f^{}\,$
will be measured at the CEPC with a typical renormalization scale
$\,\mu = M_h\,$.\, So we will input the fermion mass $\,m_f^{}\,$
as the running mass defined at $\,\mu = M_h\,$.\,
With these, we present the CEPC potential reaches (95\%\,C.L.)
in \gfig{fig:yukawaLike} and \gtab{tab:yukawaLike},
and compare them with the corresponding limits estimated for
the LHC\,(14TeV,\,300fb$^{-1}$), HL-LHC\,(14TeV, 3ab$^{-1}$), and
ILC\,(250GeV,\,250fb$^{-1}$)+(500GeV,\,500fb$^{-1}$) \cite{Peskin}.
We see that depending on the experimental precision and the involved fermion mass,
the CEPC probe of the Yukawa-type new physics scales can
reach $\,13-25\,\mbox{TeV}$ range with a $5\,\mbox{ab}^{-1}$ integrated luminosity.
These sensitivities are much higher than that of the LHC Run-2 and the HL-LHC.

From \gfig{fig:yukawaLike} and \gtab{tab:yukawaLike}, it is interesting to see that
the probe of the new physics scales with operators $\,\OO_y^\mu\,$ and $\,\OO_y^c\,$
are significantly better than other Yukawa-type operators such as
$\,(\OO_y^b,\,\OO_y^{\tau})\,$.\,
This is because the lower bound \eqref{eq:OY-limit} is proportional to
$(m_f^{}\Delta\tilde\kappa_f^{})^{-1/2}$,\, which depends on both the
coupling precision $\,\Delta\tilde\kappa_f^{}\,$ and the fermion mass $\,m_f^{}\,$.\,
Here we have used the running masses\,\cite{mass-run}\cite{PDG2014},
$(m_b^{},\,m_c^{},\,m_\tau^{},\,m_\mu^{})
 \simeq (2.41,\,0.515,\,1.713,\,0.0996)$GeV,
at the scale $\,\mu = M_h^{}\,$.\,
As shown in \gtab{tab:9} and \gfig{fig:5},  among the sensitivities to
$(\tilde\kappa_b^{},\,\tilde\kappa_c^{},\,\tilde\kappa_\tau^{},\,\tilde\kappa_\mu^{})$,\,
CEPC can measure $\,\tilde\kappa_b^{}\,$ most precisely (down to a relative precision
$\Delta\tilde\kappa_b^{}=1.27\%$), and probe $\,\tilde\kappa_\mu^{}\,$ least precisely
(down to $\Delta\tilde\kappa_\mu^{}=8.59\%$), which differ by a factor 6.76.\,
But, their running masses differ by a much larger ratio
$\,m_b^{}/m_\mu^{}\simeq\,24.2$\,.\,
This means that the fermion mass ratio has larger effect
than the ratio of their coupling sensitivities. Hence, we find that
the reach of new physics scale with $\,\OO_y^\mu\,$ is higher than that
with $\,\OO_y^b\,$ by a factor of $\,\sqrt{24.2/6.76}\simeq 1.9$\,.\,
This explains our findings shown in \gtab{tab:10}\, and \gfig{fig:6}\,.
Similarly, for the other two operators $\,(\OO_y^c,\,\OO_y^{\tau})\,$
with fermions \,$(c,\,\tau)$,\, we can deduce that the corresponding
reaches of new physics scales are enhanced relative to that of $\,\OO_y^b\,$
by a factor about $(1.8,\,1.2)$.\,

%

\vspace*{3mm}
\section{Conclusions}
\label{sec:6}
\label{sec:conclusion}
\vspace*{1mm}

The LHC Higgs discovery in 2012 has led particle physics to a turning point,
at which the precision Higgs measurements have become an important task
for seeking clues to the new physics discovery.
A future Higgs factory (like the proposed $e^+e^-$ colliders CEPC, FCC-ee, and ILC)
can provide such precision Higgs measurements.

In this work, we studied the new physics scales that a future Higgs factory can probe
via general dimension-6 operators involving the observed Higgs boson (\gtab{tab:O}).
Our analysis utilizes the existing electroweak precision observables (EWPO),
as well as the Higgs observables and precision measurements at the future
$e^+e^-$ Higgs factory (taking the CEPC as an illustration).
The conventional scheme-dependent analysis usually
fixes the three electroweak parameters ($g$, $g'$, $v$) with three high precision
electroweak observables ($\alpha, \GF, \MZ$) in the $Z$-scheme or
($\alpha, \MW, \MZ$) in the $W$-scheme, while ignoring their experimental uncertainties.
In contrast, we developed a scheme-independent approach
to incorporate full experimental information (including both central values and uncertainties)
of the EWPO in \Ssec{sec:2}. With this approach,
the electroweak parameters and the new physics scales of
dimension-6 operators can be fitted {\it simultaneously} by the same $\chi^2$ function.

The advantage of our scheme-independent approach is made clear
when the precisions of $Z$ and $W$ mass measurements become comparable at
the Higgs factory (cf.\ \gtab{tab:ZWmass-effects}). Since new physics deviations
from the SM are fairly small, as already constrained by the LHC data, the analytical
expansion up to their linear order holds well (\Ssec{sec:3}).
Accordingly, we performed the analytic linear $\chi^2$ fit
in Appendix\,\ref{sec:chi2}, which is physically intuitive,
numerically fast, and can be straightforwardly
generalized to include any number of observables and fitting parameters
under consideration.
In \Ssec{sec:4},  we demonstrated that
including the existing EWPO together with future Higgs measurements can probe
the new physics scales up to 10\,TeV (and to 44\,TeV for the gluon-involved operator
$\mathcal{O}_g^{}$) at 95\%\,C.L., as shown in \gfig{fig:NP-scale} and \gtab{tab:scale}.
We found that including the CEPC precision measurements can further lift
the reach up to 35\,TeV (\gfig{fig:4} and \gtab{tab:Zpole-effects}).
In addition, the CEPC precision tests of Higgs couplings can probe the
new physics scales with Yukawa-type operators up to $(13-25)$\,TeV,
(\gfig{fig:yukawaLike} and \gtab{tab:yukawaLike}).
We note that these indirect new physics reaches do cover the energy range to be probed by
the future hadron colliders of $pp(50-100)$\,TeV \cite{CEPC,FCC-ee}
running in the same circular tunnel.
Hence, the precision probe at the Higgs factory can provide an important guideline
for the future new physics discoveries at the SPPC or FCC-hh.
Our study demonstrates that the Higgs factory can probe the new physics of the Higgs sector
much more sensitively than what the LHC would achieve \cite{LHCeff}.

The $Z$-pole running of the $e^+e^-$ collider is required by the machine calibration at
its initial stage. In \Ssec{sec:4.6}, we further demonstrated that during the CEPC early phase,
the $Z$-pole measurements can provide even stronger indirect probe of
the new physics scales than the Higgs observables alone as measured at the Higgs factory\,($250$GeV).
This motivates a longer $Z$-pole running to ensure the no-lose probe of
new physics deviations from the SM, complementary to the Higgs factory via
the $Zh$ production.

\addcontentsline{toc}{section}{Acknowledgments\,}
\section*{Acknowledgments}
\vspace*{-2mm}

We thank Matthew McCullough, Manqi Ruan, and Tevong You for many valuable discussions.
We are grateful to Michael Peskin
for discussing the $\chi^2$ analysis of Ref.\,\cite{Peskin}.
We also thank Timothy Barklow, Tao Han, Zhijun Liang, Matthew Strassler, Liantao Wang,
Xinchou Lou and Yifang Wang for discussions.
SFG thanks Matthew McCullough for kind invitation and hospitality during his visit
at CERN Theory Division. This work is supported in part by National NSF of China (under
grants 11275101 and 11135003) and by Tsinghua University (under grant 20141081211).

\vspace*{7mm}
\noindent
{\large\bf Appendix}
\vspace*{-3mm}
\begin{appendix}

\section{Kinetic Mixing of Gauge Bosons}
\label{sec:kinetic}

The dimension-6 operators can make nontrivial corrections to both the mass matrices
and the kinetic terms of gauge bosons. The situation is much simpler for charged weak
bosons $W^\pm$ which have only one mass eigenstate and thus no extra mixing
in the effective theory of the SM with dimension-6 operators. On the other hand, the situation
for neutral gauge bosons are more involved since mixing between the photon $A$ and
the $Z$ boson can arise from either loop corrections or new physics beyond the SM. Both
kinetic mixing and mass diagonalization may appear. It is necessary to first transform
their kinetic terms into the canonical forms and mass matrix into the diagonal form before
deriving Feynman rules and computing physical processes. Here, we provide a general
formalism within this effective theory, and describe how to deal with the
corrections from dimension-6 operators up to the linear order.

\vspace*{2mm}
\subsection{Charged Gauge Bosons}
\vspace*{1.5mm}

For charged gauge boson, there is no mixing in either kinetic or mass term,
\beqa
  {\bf D}(q^2) \,=~
  \frac 1 {\,q^2 K - \left[ (M^{(0)}_W)^2 - \delta M^2_W \right]\,} \,,
\label{eq:DW}
\eeqa
where $\,K \equiv 1 + \delta K = Z_W^{-2}\,$ with $\,Z_W\,$ given by Eq.\eqref{eq:W-redef}. 
Thus, we have 
\beqa
\delta K \,\simeq\, -2\delta Z_W \,=\, -\frac {\,2v^2}{\,\Lambda^2\,} g^2 c_{WW}^{}\,. 
\eeqa
The propagator \eqref{eq:DW} reduces to its canonical form,
$
{\bf D}(q^2) = \frac 1 {\,q^2 - M^2_W\,}
$,
when the $W$ boson field and its mass are renormalized as
\begin{equation}
  W \rightarrow \frac W {\sqrt{K\,}\,}\, ,
\qquad
  M^2_W
=
  \left(\! M^{(0)}_W \right)^{\!2}
\!\left(\! 1 + \frac {\delta M^2_W}{M^2_W} - \delta K
\!\right) \!.
\label{eq:w-redef-m2}
\end{equation}
Here we have omitted the small difference between $\MW$ and $M^{(0)}_W$
in the denominator of the second term, which is of the higher order.

\vspace*{2mm}
\subsection{Neutral Gauge Bosons}
\label{sec:kinetic-NC}
\vspace*{1.5mm}

For neutral gauge bosons, both $A$ and $Z$ are involved. The kinetic mixing and mass
terms are hence $2 \!\times\! 2$ matrices. Nontrivial mixing effects can appear in both parts.
We first generally parametrize the correlated propagator as
\begin{equation}
  {\bf D}(q^2) \,=\,
  \frac 1 {\,q^2 \mathbb K - \mathbb M^2\,} \,,
\label{eq:propagator-0}
\end{equation}
where both kinetic coefficient matrix $\mathbb K$ and the mass matrix $\mathbb M$ need
to be diagonalized, $S \mathbb K S^{-1} \equiv \mathbb T$ and
$R \mathbb M^2 R^{-1} \equiv \mathbb D^2$,
with $\,\mathbb T\,$ and \,$\mathbb D$\, denoting the diagonal kinetic matrix
and diagonal mass matrix, respectively.
Then, we can first diagonalize the kinetic term as
\begin{eqnarray}
{\bf D}(q^2) \,=\,
\frac 1 {\,q^2 S^{-1} \mathbb T S - \mathbb M^2\,}
\,=\, S^{-1} \mathbb T^{- \frac 1 2}
\frac{1}{\,q^2 \mathbb I - \mathbb T^{- \frac 1 2} S
\mathbb M^2 S^{-1} \mathbb T^{- \frac 1 2}\,}
\mathbb T^{- \frac 1 2} S \,,~~~~~
\end{eqnarray}
by folding kinetic mixing to the mass matrix,
$\,\widetilde{\mathbb M}^2 \equiv
\mathbb T^{- \frac 1 2} S \mathbb M^2 S^{-1} \mathbb T^{- \frac 1 2}
$.
The modified mass matrix can be diagonalized by
$
\,\widetilde{\mathbb M}^2 \,=\,
  \widetilde R^{-1} \widetilde{\mathbb D}^2 \widetilde R\,
$.\,
Then, the propagator can be reduced to the fully diagonalized form,
\begin{equation}
  {\bf D}(q^2) \,=\,
  S^{-1} \mathbb T^{- \frac 1 2} \widetilde R^{-1}
  \frac{1}{\,q^2 \mathbb I - \widetilde{\mathbb D}^2\,} \widetilde R \mathbb T^{- \frac 1 2} S
\,\equiv\,
  \widetilde S^{T} \frac{1}{\,q^2 \mathbb I - \widetilde{\mathbb D}^2\,} \widetilde S \,.
\end{equation}

For the current effective theory, the original mass matrix
$\,\mathbb M^2 = \diag\{0, M^2_Z\}$\, of neutral gauge bosons
is diagonalized in the $A$-$Z$ space. It should be noted that
$\,M^2_Z \equiv (M^{(0)}_Z)^2 + \delta M^2_Z$\, is already the value after including
the dimension-6 operator contributions. It remains diagonalized because of the
unbroken $U(1)_{\text{em}}^{}$ gauge symmetry requires,
$\,\Pi_{AA}(0) = \Pi_{AZ}(0) = 0$,\,
when writing down the effective operators.
On the other hand, for generality, kinetic mixing can be parametrized as,
$\mathbb K \equiv \mathbb I + \delta K$,\, where $\delta K$ is a $2 \times 2$ symmetric matrix
whose explicit form will be given at the end of this section. A general feature
is that its matrix elements \,$(\delta K_{11}^{}, \delta K_{12}^{}, \delta K_{22}^{})$\, belong to
the linear order in terms of dimension-6 operator coefficients. This leads to
a sizable mixing of order $\mathcal O(1)$,
\begin{equation}
  S \,\equiv
\left\lgroup
\begin{matrix}
  \cos \theta & \sin \theta \\
- \sin \theta~\, & \cos \theta
\end{matrix}
\right\rgroup \!\!,
\quad \quad
  \tan 2 \theta
=
  \frac{2 \delta K_{12}^{}}
		{\,\delta K_{11}^{} - \delta K_{22}^{}\,} ,
\end{equation}
under which the kinetic term becomes diagonal.
But, the deviations from canonical form are still of the linear order,
$\,\mathbb T \equiv \diag \{1 + \delta K_1^{}, 1 + \delta K_2^{} \}$,\,
which is a diagonal matrix.
Then, the modified mass matrix $\,\widetilde{\mathbb M}^2$\, of neutral gauge bosons
becomes,
\begin{equation}
  \widetilde{\mathbb M}^2
\,\equiv\,
  \mathbb T^{- \frac 1 2} S \mathbb M^2 S^{-1} \mathbb T^{- \frac 1 2}
\,=\,
  M^2_Z\!
\left\lgroup
\begin{matrix}
  \frac {\sin^2 \theta}{1 + \delta K_1}
& \frac {\cos \theta \sin \theta}{\sqrt{1 + \delta K_1} \sqrt{1 + \delta K_2}}
\\[2mm]
  \frac {\cos \theta \sin \theta}{\sqrt{1 + \delta K_1} \sqrt{1 + \delta K_2}}
& \frac {\cos^2 \theta}{1 + \delta K_2}
\end{matrix}
\right\rgroup \!.~~~~~
\end{equation}
Since the rank of $\widetilde{\mathbb M}^2$ equals 1, it contains a massless
eigenstate as the photon. This property is a consequence of the unbroken
$U(1)_{\text{em}}^{}$ gauge symmetry.
In addition, the $Z$ boson mass is modified as
\begin{equation}
  \widetilde M^2_Z
\,\simeq\,
  M^2_Z
\left(
  1
\!- \sin^2 \!\theta \delta\, K_1^{}
\!- \cos^2 \!\theta \delta\, K_2^{}
\right)
\,=\,
  \left[ M^{(0)}_Z \right]^2
\!\!\left(\!
  1
+ \frac {\delta M^2_Z}{M^2_Z}
- \delta K_{22}^{}\!
\right) \!.
\end{equation}
For convenience, we have denoted the zeroth-order of $Z$ boson mass as $M^{(0)}_Z$ and
the correction as $\,\delta M^2_Z\,$ which is independent of the correction from kinetic mixing.
The modified mixing matrix $\,\widetilde R\,$ is
\begin{equation}
  \widetilde R
\,=
\left\lgroup
\begin{matrix}
  \cot\!\theta\, \sqrt{\frac {1 + \delta K_1}{1 + \delta K_2 + \cot^2 \theta (1 + \delta K_1)}}~~
& - \sqrt{\frac{1 + \delta K_2}{1 + \delta K_2 + \cot^2 \theta (1 + \delta K_1)}}
\\[4mm]
  \tan\!\theta\, \sqrt{\frac {1 + \delta K_2}{1 + \delta K_1 + \tan^2 \theta (1 + \delta K_2)}}~~
& \sqrt{\frac {1 + \delta K_1}{1 + \delta K_1 + \tan^2 \theta (1 + \delta K_2)}}
\end{matrix}
\,\right\rgroup \!\!.
\end{equation}
Altogether, we can derive the full current rotation $\,\widetilde S$\,,
\begin{equation}
  \widetilde S
\,\equiv\,
  \widetilde R \mathbb T^{- \frac 1 2} S
\,=\,
  \mathbb I -
  \frac{1}{2} \!
\left\lgroup
\begin{matrix}
  \delta K_{11}^{}\, & \,0 \\[1mm]
  2\delta K_{12}^{}\, & \,\delta K_{22}^{}
\end{matrix}
\right\rgroup \!.
\end{equation}
The mixing matrix $\,\widetilde S\,$ rotates $A$ and $Z$ as well as their
corresponding currents,
\begin{equation}
  \widetilde J
\,\equiv\,
  \widetilde S J
\,=
\left\lgroup
\begin{matrix}
  \left( 1 \!-\! \frac 1 2 \delta K_{11}^{} \right)\! J_A^{} & 0
\\[1.5mm]
0 & ~-\delta K_{12}^{} J_A^{} \!+\!
  \left(1 \!-\! \frac 1 2 \delta K_{22}^{} \right)\! J_Z^{}
\end{matrix}
\right\rgroup \!.
\end{equation}
Note that this result still has linear dependence on the dimension-6 operator coefficients.
The corrections to the kinetic term can lead to not only the field redefinitions
of $A$ and $Z$, but also the mixing between them.

For the effective operators under consideration, the $\,\delta K\,$ matrix elements are
\begin{subequations}
\begin{eqnarray}
  \delta K_{11}^{}
& = &
  \frac {-2 v^2}{ \Lambda^2}
  \left[ s^2_w g^2 c_{WW}^{} \!-\! c_w^{} s_w^{} g g' c_{WB}^{} \!+\! c^2_w g'^2 c_{BB}^{} \right] \!,
\\
  \delta K_{12}^{}
& = &
  \frac {-2 v^2}{ \Lambda^2}
  \left[ c_w^{} s_w^{} g^2 c_{WW}^{} \!-\! \frac 1 2
  (c^2_w \!-\! s^2_w) g g' c_{WB}^{} \!-\! c_w^{} s_w^{} g'^2 c_{BB}^{} \right] \!,
\\
  \delta K_{22}^{}
& = &
  \frac {-2 v^2}{ \Lambda^2}
  \left[ c^2_w g^2 c_{WW}^{} \!+\! c_w^{} s_w^{} g g' c_{WB}^{} \!+\! s^2_w g'^2 c_{BB}^{} \right] \!,
\end{eqnarray}
\label{eq:dK}
\end{subequations}
which involve only three operators, $\mathcal O_{WW}^{}$, $\mathcal O_{BB}^{}$,
and $\mathcal O_{WB}^{}$\,.

\vspace*{2mm}
\section{Analytic Linear $\chi^2$ Fit}
\label{sec:chi2}
\label{app:B}
\vspace*{1.5mm}

To make our analysis fully transparent, in this Appendix
we present the $\chi^2$ fitting method used for the current study.
With a set of observables $\mathcal O_j^{}$ to constrain model parameters,
we need to minimize the $\chi^2$ function,
\begin{equation}
  \chi^2 ~=\,
  \sum_j \left(\frac{\mathcal O^{\th}_j - \mathcal O^{\exp}_j}
  {\Delta \mathcal O_j^{}} \right)^{\!2} ,
\label{eq:chi2-exp}
\end{equation}
which is a summation of individual constraints.
In the above, we use $\mathcal O^{\text{th}}_j$ to denote the theoretical prediction,
$\mathcal O^{\exp}_j$ the experimental measurement, and $\,\Delta \mathcal O_j^{}\,$ the
associated uncertainty. The theoretical prediction is a function of model parameters.
Here, we will just use the  $\,\kappa_j^{}\,$ rescaling of
the Higgs coupling with the SM particles to fit
experimental data (as to be elaborated in \gapp{app:cepc}), for an illustration.
The deviation from the SM is then parametrized as
$\,\delta \kappa_j^{} \equiv \kappa_j^{} - 1\,$,\,
which are small numbers. When expanded to the linear term of
$\,\delta \kappa_j^{}$,\, the $\chi^2$
function can be expressed as a quadratic function with matrix manipulations,
\begin{equation}
  \chi^2 \,=\, 
  (\mathcal O^{\text{th,0}}\! + A \delta \kappa - \mathcal O^{\text{exp}})^T\,
  \overline \Sigma^{-1}
  (\mathcal O^{\text{th,0}}\! + A \delta \kappa - \mathcal O^{\text{exp}}) \,.
\end{equation}
Note that, in matrix notations, the observable
$\,\mathcal O\,$ has dimension $\,m \!\times\! 1$,\,
the deviation $\delta \kappa$ has dimension $\,n \!\times\! 1$,\,
coefficient matrix $A$ has dimension $\,m \!\times\! n$,\,
and error matrix $\,\overline \Sigma\,$ dimension
$\,m \!\times\! m$,\,
where $m$ and $n$ are the number of observables and model/fitting parameters, respectively.
The error matrix $\,\overline \Sigma^{-1}\,$ of independent measurements is diagonal,
\begin{equation}
  \overline \Sigma^{-1}
=\,
  \mbox{diag} \left\{ \frac 1 {(\Delta \mathcal O_1)^2},\, 
  \frac 1 {(\Delta \mathcal O_2)^2},\, \cdots\!,\, 
  \frac 1 {(\Delta \mathcal O_n)^2} \right\} ,
\end{equation}
according to the definition in \geqn{eq:chi2-exp}. This corresponds to uncorrelated/independent
measurements. Nevertheless, this assumption is not necessary. For correlated/dependent measurements,
the error matrix $\,\overline \Sigma^{-1}$\, in the observable basis is in general a symmetric
matrix,
$
  \overline \Sigma_{ij}^{}
\equiv
  \sigma_i^{} \rho_{ij}^{} \sigma_j^{}
$,\,
where $\,\rho\,$ is the so-called correlation matrix.

The $\chi^2$ function reaches its minimum under the condition,
$\,\partial \chi^2 / \partial \delta \kappa_j^{} = 0$\,.\,
From this, we can solve the best fit values of $\delta \kappa$,
\begin{equation}
  \delta \kappa_{\text{best}}^{}
\,=~
  (A^T \,\overline \Sigma^{-1}\! A)^{-1} A^T \,\overline \Sigma^{-1}
  \!\(\mathcal O^{\text{exp}}\! - \mathcal O^{\text{th,0}}\) .
\end{equation}
For convenience, let us rewrite the $\chi^2$ function in the fitting parameter basis,
\begin{equation}
  \chi^2
\,=~
  \chi^2_{\text{min}}
+ (\delta \kappa - \delta \kappa_{\text{best}}^{})^T \Sigma^{-1}
  (\delta \kappa - \delta \kappa_{\text{best}}^{}) ,
\label{eq:chi2-coupling}
\end{equation}
where the error matrix $\,\Sigma \equiv A^T\, \overline \Sigma^{-1} A\,$
can be obtained from $\,\overline \Sigma\,$ through matrix manipulation,
and has dimension $\,n \times n$.\,
Note that the error matrix $\,\Sigma^{-1}\,$ is also symmetric.
The $\chi^2_{\text{min}}$ can also be expressed analytically,
$\,
  \chi^2_{\text{min}}
=~
  (\mathcal O^{\text{exp}} \!- \mathcal O^{\text{th,0}})^T B^T
  \overline \Sigma^{-1} B (\mathcal O^{\text{exp}} \!- \mathcal O^{\text{th,0}})
$,\,
where
$\,B \equiv \mathbb I - A (A^T \,\overline \Sigma^{-1} A)^{-1} A^T\, \overline \Sigma^{-1}$.\,
If the theoretical prediction is consistent with experimental measurement,
$\,\mathcal O^{\th,0}_j \!= \mathcal O^{\exp}_j$,\, the $\chi^2$ function reaches the minimum
at the SM values, $\,\delta \kappa_j^{} = 0$,\, which is the best value. This formalism of
analytic $\chi^2$ function can even be used to estimate the statistical fluctuation
in $\chi^2_{\min}$ \cite{Ge:2012wj}.

In general, different fitting parameters are correlated with each other through the coefficient
matrix $A$ and hence can affect each other. To obtain the precision of a specific fitting parameter,
we need to marginalize over the others. This can be done as a series of iterative reductions from
higher-dimensional $\chi^2$ function to lower one, each time reducing the number of fitting parameters
by $1$. During this process, the $\chi^2$ function can
still be expressed with the quadratic form \geqn{eq:chi2-coupling} in the fitting parameter basis while
the n-dimensional error matrix $\,\Sigma\,$ can be reduced to $(n-1)$-dimensional
$\,\widetilde \Sigma\,$
by integrating out one degree of freedom, say the $k$-th branching fraction,
\begin{equation}
  \widetilde \Sigma^{-1}_{ij}
~=~
  \Sigma^{-1}_{ij}
- \frac {\,\Sigma^{-1}_{ik} \Sigma^{-1}_{jk}\,}{\Sigma^{-1}_{kk}} .
\label{eq:reduction}
\end{equation}
Note that there is no summation over $k$. This reduction formula is just a reflection of
integrating out the $k$-th degree of freedom from the
probability distribution $\,\mathbb P(\delta \kappa_j^{}) \equiv \exp(- \chi^2/2)$,
\beqa
\mathbb P(\delta \kappa_1^{} \cdots \delta \hat{\kappa}_k^{} \cdots \delta \kappa_n^{})
~=\,  \int^{+\infty}_{-\infty}
\mathbb P (\delta \kappa_1^{} \cdots \delta \kappa_k^{} \cdots \delta \kappa_n^{})
\,\text{d} \delta \kappa_k^{}
\,.
\eeqa
The hat means that the corresponding variable has been integrated out. With quadratic $\chi^2$,
this is an integration of Gaussian distribution that can be done analytically to produce
\geqn{eq:reduction}. The same procedure should be carried out until there is only one degree
of freedom left, say, the $\ell$-th anomalous coupling. The only element
of the 1-dimensional error matrix is then its uncertainty,
$\,\Delta(\delta \kappa_{\ell}^{}) \equiv
  \sqrt{\widetilde \Sigma_{\ell\ell}^{}} $\,.\,
To deduce the precision of all fitting parameters, we need to run over all possible values of
$\,\ell\,$ and make the reduction for each case. This analytic $\chi^2$ fitting technique,
along with other extensions, will be delivered in a general purpose package
BSMfitter \cite{BSMfitter}.

\end{appendix}


\end{document}